\documentclass[global,twocolumn]{svjour}
\usepackage{latexsym}
\usepackage{graphicx}
\journalname{myjournal}
\begin{document}
\titlerunning{Production of Radioisotopes in Photonuclear Reactions} 
\title{Production of Medical Radioisotopes with High Specific Activity 
in Photonuclear Reactions with $\gamma$ Beams of High Intensity 
and Large Brilliance}
\author{D.~Habs\inst{1}, 
and U.~K\"oster\inst{2}} 
%
\offprints{}          
\institute{$^1$ Fakult\"at f\"ur Physik, 
       Ludwig Maximilians Universit\"at M\"unchen, D-85748 Garching, Germany\\
           $^2$ Institut Laue Langevin, 6 rue Jules Horowitz, 
           F-38042 Grenoble Cedex 9, France}
\date{Received: date / Revised version: date}
\maketitle
\begin{abstract}
We study the production of radioisotopes for nuclear medicine 
in $(\gamma,x{\rm n}+y{\rm p})$ photonuclear reactions or 
($\gamma,\gamma'$) photoexcitation reactions
with high flux [($10^{13}-10^{15}$)$\gamma$/s], small diameter 
$\sim (100 \, \mu$m$)^2$ and small band width 
($\Delta E/E \approx 10^{-3}-10^{-4}$) $\gamma$ 
beams produced by Compton back-scattering of laser light from relativistic 
brilliant electron beams. We compare them to (ion,$x$n$ + y$p) reactions 
with (ion=p,d,$\alpha$) from particle accelerators like cyclotrons and 
(n,$\gamma$) or (n,f) reactions from nuclear reactors. 
For photonuclear reactions with a narrow $\gamma$ beam the 
energy deposition in the target can be managed
by using a stack of thin target foils or wires, hence avoiding 
direct stopping of the Compton and pair electrons (positrons).
However, for ions with a strong atomic stopping only a fraction of less 
than $10^{-2}$ leads to nuclear reactions resulting in a target heating, which 
is at least $10^{5}$ times larger per produced radioactive ion 
and is often limits the achievable 
activity. In photonuclear reactions the well defined 
initial excitation energy of the compound nucleus leads to a small 
number of reaction channels and enables new combinations of target
isotope and final radioisotope. The narrow bandwidth $\gamma$ 
excitation may make use of the fine structure of the Pygmy Dipole Resonance
(PDR) or fluctuations in $\gamma$-width leading to increased cross sections. 
Within a rather short period compared to the isotopic half-life, a target 
area of the order of $(100 \,\mu$m$)^2$ can be highly 
transmuted, resulting in a very high specific activity. $(\gamma,\gamma')$ 
isomer production via specially selected $\gamma$ cascades allows to produce 
high specific activity in multiple excitations, where no back-pumping of 
the isomer to the ground state occurs. We discuss in detail many specific 
radioisotopes for diagnostics and therapy applications.
Photonuclear reactions with $\gamma$ beams allow to 
produce certain radioisotopes, e.g. $^{47}$Sc, $^{44}$Ti, $^{67}$Cu, 
$^{103}$Pd, $^{117m}$Sn, $^{169}$Er, $^{195m}$Pt or $^{225}$Ac,
with higher specific activity and/or more economically than 
with classical methods. This will open the way for completely 
new clinical applications of radioisotopes. For example $^{195m}$Pt 
could be used to verify the patient's response to chemotherapy 
with platinum compounds before a complete treatment is performed.
Also innovative isotopes like $^{47}$Sc, $^{67}$Cu and $^{225}$Ac could be produced 
for the first time in sufficient quantities for large-scale application in 
targeted radionuclide therapy.
\end{abstract}

\sloppy
\section{Introduction}
In nuclear medicine radioisotopes are used for diagnostic and 
therapeutic purposes \cite{schiepers06,cook06}. Many diagnostics 
applications are based on molecular imaging methods, i.e. either 
on positron emitters for 3D imaging with PET (positron emission 
tomography) or gamma ray emitters for 2D imaging with planar 
gamma cameras or 3D imaging with SPECT (single photon emission 
computer tomography)\footnote{Today the nuclear medicine imaging 
techniques PET and SPECT (using radionuclides injected into 
the patient's body) are frequently combined in the same apparatus 
with the radiology technique CT (computer tomography) to PET/CT 
or SPECT/CT, respectively. CT is based on transmission radiography 
with X-rays and provides structural information to better localize 
the features observed on PET or SPECT images. CT information helps 
also to perform the attenuation correction. However, the combination 
with CT, or not, does not affect the choice of the PET or SPECT tracer 
isotopes which is the main issue in the present discussion. Hence we 
use PET and SPECT, which implicitly includes CT where adequate.}. 
The main advantage of nuclear medicine methods is the
high sensitivity of the detection systems that allows using tracers at
extremely low concentrations (some pmol in total, injected in typical
concentrations of nmol/l). This extremely low amount of radiotracers assures
that they do not show any (bio-)chemical effect on the organism. Thus, the
diagnostic procedure does not interfere with the normal body functions and
provides direct information on the normal body function, not perturbed 
by the detection method. Moreover, even elements that would be chemically 
toxic in much higher concentrations can be safely used as 
radiotracers (e.g. thallium, arsenic, etc.). To maintain these 
intrinsic advantages of nuclear medicine diagnostics
one has to assure that radiotracers of relatively high 
specific activity are used, i.e. that the injected radiotracer 
is not accompanied by too much stable isotopes of the same 
(or a chemically similar) element.

Radioisotopes are also used for therapeutic applications, in 
particular for endo-radiotherapy. Targeted systemic therapies 
allow fighting diseases that are non-localized, e.g. leukemia and 
other cancer types in an advanced state, when already multiple 
metastases have been created. Usually a bioconjugate 
\cite{schiepers06} is used that shows a 
high affinity and selectivity to bind to peptide receptors 
or antigens that are overexpressed 
on certain cancer cells with respect to normal cells. 
Combining such a bioconjugate with a suitable radioisotope such as a
(low-energy) electron or alpha emitter, allows irradiating and destroying
selectively the cancer cells. Depending on the nature of the bioconjugate,
these therapies are called Peptide Receptor Radio Therapy (PRRT) 
\cite{cook06,Reu06} when peptides are used as bioconjugates or 
radioimmunotherapy (RIT) \cite{cook06,Jac10}, when 
antibodies are used as bioconjugates. Bioconjugates could 
also be antibody-fragments, nanoparticles, microparticles, etc.  
For cancer cells having only a limited number of selective binding sites, 
an increase of the concentration of the bioconjugates may lead to 
blocking of these sites and, hence, to a reduction in selectivity. 
Therefore the radioisotopes for labeling of the bioconjugates should 
have a high specific activity to minimize injection of bioconjugates 
labeled with stable isotopes that do not show radiotherapeutic efficiency. 
Thus often high specific activities are required for radioisotopes used 
in such therapies.


The tumor uptake of bioconjugates varies considerably from one patient 
to another. This leads to an important variation in dose delivered to 
the tumor if the same activity (or activity per body mass or activity 
per body surface) was administered. Ideally a personalized
dosimetry should be performed by first injecting a small quantity 
of the bioconjugate in question, marked by an imaging isotope 
(preferentially $\beta^+$ emitter for PET). Thus the tumor uptake
can be quantitatively determined and the injected activity of the 
therapy isotope can be adapted accordingly. To assure a representative 
in-vivo behaviour of the imaging agent, the PET tracer should be ideally 
an isotope of the same element as the therapy isotope, or, at least of 
a chemically very similar element such as neighboring lanthanides. 
Thus so-called ``matched pairs'' of diagnostic and therapy
isotopes are of particular interest: $^{44}$Sc/$^{47}$Sc, 
$^{61}$Cu or $^{64}$Cu/$^{67}$Cu, $^{86}$Y/$^{90}$Y,
$^{123}$I or $^{124}$I/$^{131}$I or $^{152}$Tb/$^{149}$Tb or $^{161}$Tb. 
Often the production 
of one of these isotopes is less straightforward with classical methods. 
Therefore ``matched pairs'' are not yet established as standard
in clinical practice. The ``matched pairs'' of scandium and copper can be
produced much better with $\gamma$ beams.
Valence-III elements do not necessarily show an 
identical in-vivo behaviour \cite{Bey00,Reu00} but in many cases they are
sufficiently similar. For example the 68~min PET tracer $^{68}$Ga is conveniently 
eluted from $^{68}$Ge generators and used as imaging analog for the 
therapy isotopes $^{90}$Y, $^{177}$Lu or $^{213}$Bi \cite{Mae05}. 

The radioisotopes for diagnostic or therapeutic nuclear medicine applications
are usually produced by nuclear reactions. The required projectiles are
typically either neutrons (from dedicated irradiation reactors) or charged
particles (from small or medium-sized cyclotrons or other accelerators). 
In section 2 we shortly discuss these presently used techniques and then
introduce in section 3 the new $\gamma$ beams with high $\gamma$ energies,
high intensities and small bandwidth. Such a $\gamma$ facility 
will typically consist of an electron linac, delivering a relativistic 
electron beam with high brilliance and high intensity from which 
intense laser beams are Compton back-scattered. These $\gamma$ facilities
allow to produce many radioisotopes in new photonuclear reactions with 
significantly higher specific activity. In section 4 we compare for 
certain radioisotopes of interest the specific activities achievable 
with presently used production reactions and $\gamma$-beams respectively.
In section 5 the energy deposition of $\gamma$ beams is compared to ion 
beams, showing that targets can endure higher intensities of $\gamma$ beams 
than ion beams allowing for higher $\gamma$ flux densities.  
We will discuss interesting cases of specific radioisotopes in section 6.  
Besides attaching radiosiotopes to biomolecules in therapeutical applications, 
we discuss in section 7 new ways of brachytherapy. Finally in section 8 
the advantages of producing radioisotopes by $\gamma$ beams are outlined.

\section{Presently used Nuclear Reactions to Produce Medical Radioisotopes}

Today the most frequently employed nuclear reactions for the production 
of medical radioisotopes are:
\begin{itemize}
\item[1] {\it Neutron capture}\\
 Neutron capture (n,$\gamma$) reactions transmute a stable isotope into a
  radioactive isotope of the same element. High specific activities are
  obtained, when the (n,$\gamma$) cross section is high 
  and the target is irradiated in a high neutron flux. Neutrons most useful 
  for (n,$\gamma$) reactions have energies from meV to keV (thermal and 
  epithermal neutrons) and are provided in the irradiation positions of 
  high flux reactors at flux densities of $10^{14}$ n/(cm$^2$s)
  up to few $10^{15}$ n/(cm$^2$s). If the neutron 
  capture cross section is sufficiently high (e.g. 2100 barn for 
  $^{176}$Lu(n,$\gamma$)$^{177}$Lu), then a good fraction of the target atoms 
  can be transmuted to the desired product isotopes, resulting in a product of 
  high specific activity.

  Note that the specific activity of the product depends on the neutron 
  flux density (n/(cm$^2$s)) and not on the total number of neutrons provided 
  by the  reactor. Hence irradiation reactors are optimized to provide a high 
  flux density in a limited volume, while keeping the total neutron rate 
  (that is proportional to the thermal power) relatively low. This 
  optimization is inverse to a power reactor that should provide a
  high thermal power at limited neutron flux density (to limit the power 
  density, damage to structural materials and extend the operation time 
  between refuelling).

High specific activities can also be achieved by using indirect 
production paths. The (n,$\gamma$) reaction is not populating directly the 
final product but a precursor that decays by beta decay to the final product.
Thus the final product differs in its chemical properties from the target 
and can be chemically separated from the bulk of the remaining target material.
This method is e.g. used to produce $^{177}$Lu in non-carrier added quality 
by irradiating enriched $^{176}$Yb targets in a high neutron flux to produce 
$^{177}$Yb that subsequently decays to $^{177}$Lu.
The latter is extracted by a chemical Lu/Yb separation. 



\item[2] {\it Nuclear fission}\\
Fission is another process used for isotope production in nuclear reactors. 
Radiochemical separation leads to radioisotopes of ``non-carrier-added'' 
quality, with specific activity close to the theoretical maximum.
Fission is the dominant production route for the generator isotopes 
$^{99}$Mo and $^{90}$Sr, for the $\beta^-$ emitting therapy isotope $^{131}$I 
and for the SPECT isotope $^{133}$Xe.

\item[3] {\it Charged particle reactions with p, d or $\alpha$ ions}\\
Imaging for diagnostic purposes requires either $\beta^+$ emitters for 
PET (mainly $^{18}$F, $^{11}$C, $^{13}$N, $^{15}$O, $^{124}$I or $^{64}$Cu), 
or isotopes emitting gamma-rays with suitable energy for SPECT (about 70 
to 300 keV), if possible without $\beta^{+/-}$ emission to minimize the 
dose to the patient. Thus electron capture decay is preferred for such 
applications, e.g.: $^{67}$Ga, $^{111}$In, $^{123}$I, $^{201}$Tl.
Usually these neutron-deficient isotopes cannot be produced by neutron capture 
on a stable isotope (exception $^{64}$Cu). Instead they are mainly 
produced by charged-particle induced reactions such as (p,n), (p,2n),\dots\ 
High specific activities of the final product are achievable, when the 
product differs in chemical properties from the target (i.e. different $Z$) 
and can be chemically separated from the remaining bulk of target 
material\footnote{Note that in principle very high specific activities may 
be achieved in this way. However, the effective specific activities that 
measure the ratio between wanted radioisotopes to all elements 
(mostly metals) that may affect, e.g., the labeling of bioconjugates, might 
be significantly lower since stable elements can be introduced due to 
the finite purity of chemicals, columns, etc. used in the chemical processing. 
Such problems are generally reduced when higher activities per batch
are processed.}. Thus $Z$ must be changed in the nuclear reaction, e.g. 
in (p,n), (p,2n), (p,$\alpha$) reactions. The energies of the charged 
particle beams for such reactions are usually in the range of 10 to 
30 MeV and can be supplied with high currents (0.1 to 1 mA) 
by small cyclotrons.


\begin{table}[h]
\caption{Radioisotopes for nuclear medicine produced in generators. 
$^*$: $^{212}$Pb and $^{213}$Bi are the grand-grand-daughters and 
$^{212}$Bi is the grand-grand-grand-daughter
of the respective generator isotope.}
\bigskip
\begin{center}
\begin{tabular}{cccc} \hline
mother isotope   & $T_{1/2}$& daughter isotope& $T_{1/2}$\\ \hline
$^{44}$Ti        &  60.4 a  & $^{44}$Sc      &  3.9 h   \\
$^{52}$Fe        &  8.3 h   & $^{52}$Mn      &  21 m  \\
$^{68}$Ge        &  288 d   & $^{68}$Ga      &  68 m  \\
$^{81}$Rb        &  4.6 h   & $^{81}$Kr      &  13 s  \\
$^{82}$Sr        &  25.0 d  & $^{82}$Rb      &  76 s  \\  
$^{90}$Sr        &  28.5 a  & $^{90}$Y      &  64 h  \\  
$^{99}$Mo        &  66 h    & $^{99m}$Tc     &  6.0  h   \\
$^{188}$W        &  69 d  & $^{188}$Re     &  17 h   \\    
$^{224}$Ra       &  3.7 d  & $^{212}$Pb$^*$     &  10.6 h \\
$^{224}$Ra       &  3.7 d  & $^{212}$Bi$^*$     &  61 m \\
$^{225}$Ac       &  10 d  & $^{213}$Bi$^*$     &  45 m \\
\hline
\end{tabular}
\end{center}
\end{table}

\item[4] {\it Generators}\\
Another important technique is the use of generators, where
short-lived radionuclides are extracted  ``on-tap''
from longer-lived mother nuclides. Here the primary radioisotope 
(that was produced in the nuclear reaction) has a longer half-life than 
the final radioisotope (that is populated by decay of the primary 
radioisotope and is used in the medical application). The primary radioisotope
is loaded onto the generator and stays there chemically fixed. The final 
radioisotope will grow in, populated by the decay of the primary radioisotope.
It can be repetitively eluted and used. For the extraction of the 
shorter-lived isotope chromatographic techniques, distillation or 
phase partitioning are used. Depending on the generator technology, 
there is usually a limit to which a generator can be loaded with atoms 
of the primary product element (e.g. molybdenum on acid alumina columns). 
If more is loaded, then a significant part of the
primary product isotope might be eluted too (``breakthrough''), leading
to an inacceptable contamination of the product with long-lived
activity. To prevent such problems, generators are generally loaded with
material of a given minimum specific activity.

\item[5] {\it Photonuclear reactions}\\
The inverse process to (n,$\gamma$), namely ($\gamma$,n), also allows 
producing neutron deficient isotopes, but conventional $\gamma$ ray 
sources do not provide sufficient flux density for efficient production 
of radioisotopes with high total activity and high specific activity. 
Therefore this process plays no role in present radioisotope supply.
\end{itemize}

\section{\boldmath $\gamma$ Beams \unboldmath} 

\begin{figure}[t]
\centerline{\includegraphics[width=.47\textwidth]{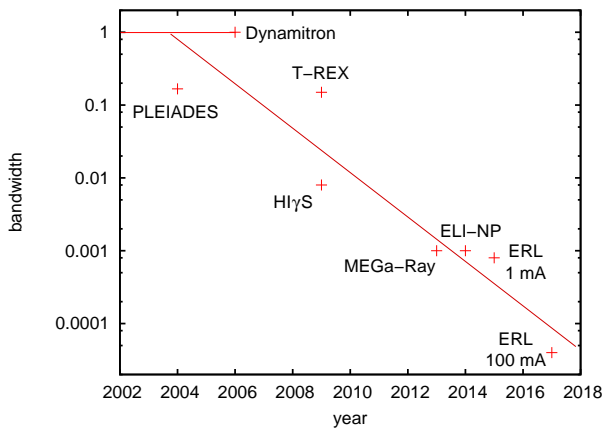}}
   \caption{Bandwidth of high energy $\gamma$ beams ($\approx$ 10 MeV)
as a function of time.}
   \label{fig1}
\end{figure}

\begin{figure}[t]
\centerline{\includegraphics[width=.47\textwidth]{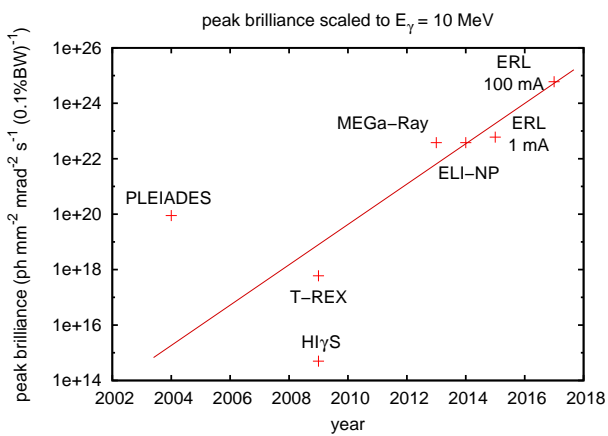}}
   \caption{Peak brilliance of high energy $\gamma$ beams ($\approx 10$ MeV)
            as a function of time.}
   \label{fig2}
\end{figure}

The new concept of isotope production with a $\gamma$ beam only became 
possible, because very brilliant $\gamma$ sources are being 
developed, where the $\gamma$ rays are produced by incoherent Compton 
back-scattering of laser light from brilliant high-energy electron bunches.
Fig.~\ref{fig1} and Fig.~\ref{fig2} show the rapid progress of 
$\gamma$ beam properties for the bandwidth (Fig.~\ref{fig1})
and the peak brilliance (Fig.~\ref{fig2}) with time, starting with the  
bremsstahlung spectrum of the Stuttgart Dynamitron \cite{kneissl06}, which
still had a very large bandwidth.

For Compton back-scattering in a head-on collision 
the $\gamma$ energy is given by:

\begin{equation}
   E_{\gamma}=\frac{4\gamma_e^2 E_L}{1+
              (\gamma_e \Theta_{\gamma})^2 + 4\gamma_e E_L/mc^2}
\end{equation}

with the $\gamma_e$ factor, characterizing the energy of the electron 
beam, the $\gamma$ energy $E_{\gamma}$, its angle $\Theta_{\gamma}$ and the
laser photon energy $E_L$. The energy $E_{\gamma}$ decreases with
$\Theta_{\gamma}$. A small bandwidth of the $\gamma$ beam requires a small
energy spread of the electron bunches $(\Delta \gamma_e/\gamma_e)$, a small
bandwidth of the laser energy $(\Delta E_L/E_L)$ , a very good emittance
of the electron beam with a small opening angle and small opening angle of the 
laser beam. At the HI$\gamma$S facility (Duke University, USA) the photons 
are produced by a Free Electron Laser (FEL) and then are back-scattered 
from a circulating electon beam \cite{weller09}. This facility already 
produced high energy $\gamma$ rays (1-100 MeV),
but the flux ($10^5 - 5 \cdot 10^8 \gamma$/s) was too weak for radioisotope production.
C.~Barty and his group at the Lawrence Livermore
National Laboratory (LLNL) developed already three generations
of incoherent Compton back-scattering sources: PLEIADES \cite{pleiades04},
T-REX \cite{trex10} and MEGa-Ray \cite{barty10}, each based on a ``warm'' 
electron linac and a fibre laser for back-scattering. Recently the electron 
linac technology was switched from S-band technology (4 GHz) for T-REX
to X-band technology (12 GHz) for MEGa-Ray. Here electron bunches
with 250 PC are used. The  MEGa-Ray
$\gamma$ beam runs with a macro-pulse structure
of 120 Hz using  1.5 J, 2 ps laser pulses, which are recirculated 100 times
with 2 ns bunch spacing in a ring-down cavity. The group plans for lower energy
$\gamma$ rays in the range of a only few MeV, 
too small for photonuclear reactions.
A similar $\gamma$ facility is planned for 
the ELI-Nuclear Physics project (ELI-NP) in Romania \cite{ELI-NP10}, 
also based on a ``warm'' linac like the one used at MEGa-Ray, however 
designed for $\gamma$ energies up to 19 MeV, thus reaching interesting 
intensities and $\gamma$ energies for isotope production. 
R. Hajima and coworkers at Ibaraki (Japan)
are developping  a Compton back-scattering 
$\gamma$ beam using an energy recovery linac (ERL) and superconducting ``cold''
cavities \cite{hajima09}. For smaller electron bunch 
charges (8 pC)  very low normalized emittances of 
0.1 mm mrad can be obtained from the electron gun. For the reflected laser 
light a high finesse enhancement cavity is used for recirculating the photons. 
The quality of the electron beam from the ERL can be preserved by running 
with higher repetition rate (GHz). Switching from a 1 mA electron current to 
a 100 mA current the peak brilliance and bandwidth can be improved 
significantly \cite{hajima10,ERL-GAMMA08}. Intensities of 
$5\cdot 10^{15}\gamma$/s are expected \cite{hajima10}.

\begin{figure*}[t!]
\centerline{\includegraphics[width=.8\textwidth]{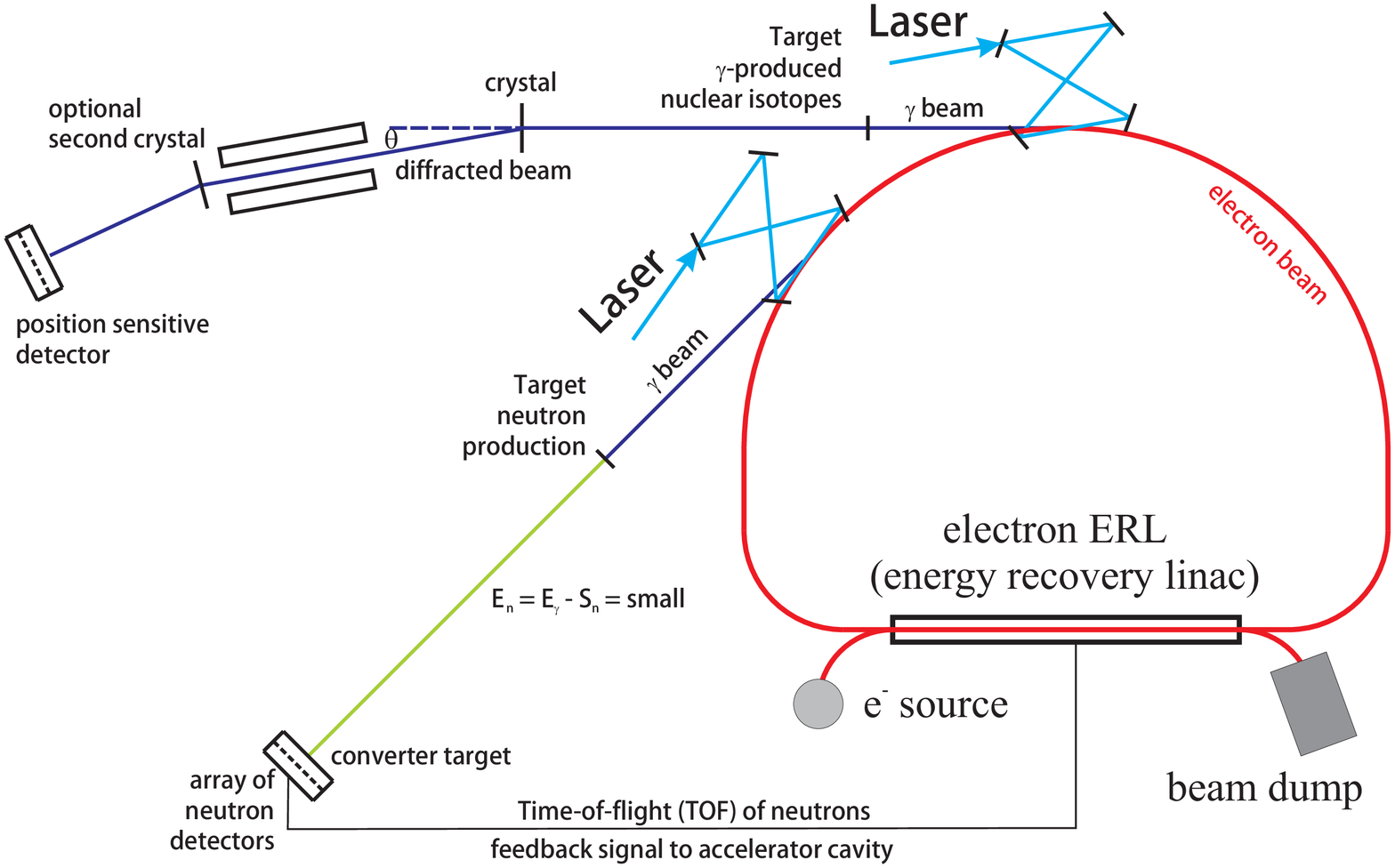}}
   \caption{ Energy Recovery Linac (ERL) electron beam facility 
with 2 $\gamma$-beam production points and a crystal spectrometer and 
neutron time of flight (TOF) spectrometer to monitor the $\gamma$ energy. 
The electron beam starts from the electron source, 
is accelerated in the superconducting cavity and then, after one loop, 
is decelerated again in the same cavity, feeding back its energy, 
and then is stopped in the beam dump. For Compton back-scattering the laser 
light is recycled many times in the enhancement cavity.}
   \label{fig3}
\end{figure*}

Also laser-accelerated electron bunches have been proposed as 
relativistic mirrors for Compton back-scattering and the production
of intense $\gamma$ beams \cite{habs08}. 

The yield of resonant photonuclear reactions (discussed below) depends 
strongly on the exact energy and the bandwidth of the gamma beam. Both 
parameters are determined by the quality of the laser beam and of the 
electron beam respectively. The laser beam parameters are usually well 
controlled by conventional means used in laser spectroscopy. More 
importantly, the electron beam parameters need to be tuned and monitored 
with good precision. For a meaningful monitoring system the 
$\gamma$ ray energy has to be measured with a system that has (far) 
better energy resolution than the $\gamma$ beam itself. It is not trivial 
to measure high energy gamma ray energies with such high precision. For 
gamma rays in the MeV range usual Ge detectors are limited to 
an energy resolution of the order of $10^{-3}$. Scintillation detectors 
have an even worse energy resolution. Hence, more complex and ``unusual'' 
methods have to be used for this purpose. The two methods are
shown schematically in Fig.~\ref{fig3}, where an energy recovery linac
can supply many $\gamma$ beams.

Two methods are preferable:
\begin{description}
\item[{\it A crystal spectrometer}] 
A thin single or mosaic crystal (e.g. Si, Ge, SiO$_2$, Cu, graphite, etc.) 
is placed in the gamma beam (in front, inside or behind a production target). 
A small fraction of the  beam will be diffracted by the crystal according 
to the Bragg condition: 
\begin{equation} 2d \sin \theta = n \lambda \end{equation}
$d$ is the well-known crystal lattice spacing, $n$ is the reflection order, 
$\lambda$ is the wavelength of 
the $\gamma$ beam and $\theta$ is the diffraction angle.
Placing a $\gamma$ ray detection system at large distance allows 
measuring the diffraction angle $\theta$, 
either by scanning the beam through narrow collimators by turning 
the crystal or by using a fixed crystal 
and a detector with good position resolution. Hence. the wavelength 
of the $\gamma$ rays is deduced which 
gives directly the $\gamma$ ray energy.
The angular spread of the diffracted beam is a measure of the energy 
spread of the $\gamma$ beam. 
These data can be used for a feedback system for tuning and monitoring 
the electron beam for the $\gamma$ beam production. 
Due to the high intensity of the $\gamma$ beam, even with thin crystals 
and in high reflection order enough photons 
will arrive at the detector. A higher reflection order is preferred 
since it allows placing the detector further away from the direct 
non-diffracted beam. For $\gamma$ beams of larger opening angle, 
the latter would limit the achievable energy resolution. Here it is 
preferable to use two consecutive crystals for diffraction: a first one 
placed in the $\gamma$ beam to diffract out a small fraction. The small 
intrinsic angular acceptance of the crystal will effectively act as 
collimator. The second crystal receives therefore a well-collimated beam. 
Additional collimators can be placed between both crystals and/or between 
the second crystal and the detector respectively to eliminate background 
from other diffraction orders. Using two consecutive diffractions in the 
same direction will add to the energy dispersion and provide very high 
energy resolution, two diffractions in opposite direction allows measuring 
the intrinsic resolution of the measurement system. The rotation angle of 
the crystals is usually controlled by laser interferometers. Such a double 
crystal spectrometer enables measuring $\gamma$ ray energies with a 
resolution below $10^{-6}$ \cite{Dew06}, i.e. covering fully the needs 
to stabilize the $\gamma$ beam in a bandwidth of $10^{-5}$ to $10^{-3}$. 
\item[{\it B ($\gamma$,n) threshold reaction with neutron time-of-flight spectro\-meter}]
Alternatively to a crystal spectrometer also a second $\gamma$ beam 
from a second $\gamma$ ray production station (possibly using a 
different laser wavelength) can be used for monitoring the electron beam 
energy. This second $\gamma$ beam is sent on a dedicated target where 
it induces ($\gamma$,n) reactions just above the threshold and neutrons 
are released in the eV to keV range. Due to the pulsed nature of the 
$\gamma$ beam, the neutron energy can be measured by time-of-flight with 
good precision (few eV or better). Adding the measured neutron energy to 
the well-known neutron binding energy of the target provides an accurate 
on-line measurement (order of $10^{-6}$ resolution) of the $\gamma$ beam 
energy and $\gamma$ beam energy spread, hence also of the electron beam 
energy and electron beam energy spread. Again this information is used 
for a feedback system to optimize and stabilize the electron accelerator 
parameters. Neutron detection can be realized in various way. One 
possibility is the use of a ``neutron converter'' (containing isotopes 
like $^6$Li, $^{10}$B or $^{235}$U) combined with a charged particle 
detector. Using a segmented detector array many neutrons could be measured 
per bunch allowing for a fast feedback system. The length of the neutron 
flight path should be adjusted to the neutron energies.
\end{description}

\begin{figure*}[t!]
\centerline{\includegraphics[width=.6\textwidth]{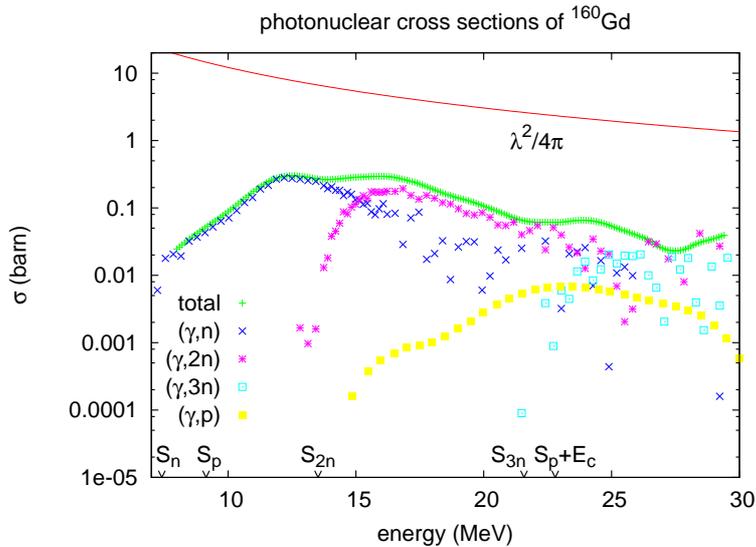}}
   \caption{ Measured photonuclear cross section for $^{160}$Gd.
Also shown is the maximum Breit-Wigner resonance cross 
section $\frac{\lambda^2}{4\pi}$ as an orange curve. The threshold values
for the reactions are indicated by arrows as well as the energy 
$S_p + E_c$ where Coulomb hindrance of proton emission disappears. 
The data are taken from Ref. \cite{data}.}
   \label{fig4}
\end{figure*}

\section{Specific Activity of Radioisotopes and Photonuclear Cross Sections}

One of the most important quality criteria for radioisotopes for
nuclear medicine applications is the specific activity (A/m), usually
expressed in GBq/mg, Ci/mg or similar units. The theoretical maximum specific
activity for a pure radioisotope without admixture of stable isotopes is
given by:

\begin{equation}
\left(\frac{A}{m}\right)_{\rm max} = \frac{\ln(2)}{T_{1/2}} \frac{\cdot N_A}{M}
\label{eq1}
\end{equation}

Where $T_{1/2}$ is the half-life of the radioactive isotope,  
 $M$ is the molar mass (g/mol) of the target isotope and $N_A=6.02\cdot10^{23}$
denotes Avogadro's constant.

For example an isotope with $M=100$~g/mol and $T_{1/2} = 7$~days would have a
theoretical specific activity of nearly 7 TBq/mg.

If other reactions (such as destruction of the product by nuclear reactions) 
do not interfere significantly, then the achievable specific activity is 
given by:

\begin{equation}
\frac{A}{m} = \frac{N_A}{M} \sigma \cdot\Phi\cdot \left[1 -
\exp(-\ln(2) t_{\rm irr}/ T_{1/2}\right)]
\label{eq2}
\end{equation}

Where:

\begin{tabular}{ll}
$M$ &		the molar mass of the target isotope\\
$\sigma$ & the cross-section for transmutation of the\\ 
         &  target into the product\\
$\Phi$ & 	the particle flux density\\
$t_{\rm irr}$&	the irradiation time\\
$T_{1/2}$ & 	the half-life of the product isotope
\end{tabular}

After sufficiently long irradiation (multiple of product half-life) the
specific activity approaches saturation:
\begin{equation}
\frac{A}{m} = \frac{N_A}{M} \sigma \cdot \Phi
\label{eq3}
\end{equation}

Comparison of equations \ref{eq1} and \ref{eq3} shows that the necessary
condition to reach high specific activities is: 
$\sigma\cdot\Phi \approx \ln(2)/T_{1/2}$

Radioisotopes for medical applications have typically half-lives of hours to
days, hence the flux density (in part./(cm$^2$s)) should approach or exceed 
a value of about $10^{19}$/$\sigma$ (in barn). For future planned $\gamma$ 
beams with $5\cdot 10^{15}\gamma$/s over areas of (0.1 mm)$^2$ the flux 
density can reach several $10^{19}\gamma$/(cm$^2$ s), i.e. the target can 
be efficiently transmuted by photonuclear reactions with few 100~mb 
cross-section. For resonant reactions with higher cross-sections even 
the flux densities of the less powerful $\gamma$ beam facilities 
($10^{17}\gamma$/(cm$^2$ s)) will assure a relatively
high specific activity of the product.

The specific activity of a radioisotope product in units of GBq/mg or 
Ci/mg is a measure of quality familiar to the users in nuclear medicine. 
However, the theoretical specific activity varies with the
half-life. Hence a tabulation of different isotopes and their achievable 
specific activities is less comprehensible. Therefore we define in addition 
the ratio $R=$[(A/m)/(A/m)$_{\rm max}$] that indicates how close the 
specific activity comes to the theoretical optimum. For $R=0.5$ one 
radioactive atom will be accompanied by one stable atom (of the same 
element or target, respectively), for $R=0.1$ only one out of ten atoms 
is the radioisotope of interest, etc.

We will compare 
$R_{({\rm n},\gamma)}$=[(A/m)/(A/m)$_{\rm max}$]$_{({\rm n},\gamma)}$ 
for classical production in (n,$\gamma$) reactions to \\ $R_{\gamma}$=
[(A/m)/(A/m)$_{\rm max}$]$_{\gamma}$ for $\gamma$-beams.

The finally reached specific activity is also determined by the 
undesired further transmutation (burnup) of the wanted reaction product. 
This product burnup becomes significant when the product fraction gets high.
For (n,$\gamma$) reactions in high flux reactors it may eventually limit 
the achievable specific activity if the neutron capture cross-section of 
the product is high. For $^{153}$Gd, $^{159}$Dy, $^{169}$Yb or $^{195m}$Pt
this seriously limits the achievable specific activity. 
For photonuclear reactions the cross sections for product creation 
and destruction are comparable in case of ($\gamma$,n) reactions, while 
for ($\gamma$,p) and ($\gamma$,2n) reactions the destruction cross-section 
may be up to one order of magnitude larger. This limits the ultimately 
achievable specific activity to $R \approx 0.5$ in the first case 
and to $R \approx 0.1$ in the latter cases. Product burn\-up becomes 
noticeable when approaching these limits. If the cross-sections were known, 
the effect could be calculated precisely by the Bateman equations. 

In some cases a secondary product produced by a reaction on the primary 
desired product may present a disturbing radionuclide impurity. For example 
when $^{125}$I is produced by $^{124}$Xe(n,$\gamma$)$^{125}$Xe($\beta^-$), 
then the irradiation should be kept short enough to minimize production 
of disturbing $^{126}$I by $^{125}$I(n,$\gamma$) reactions.

If one looks at measured photonuclear cross-sections one typically finds
cross sections below 1 barn. As a prototype we show in 
Fig.~\ref{fig4} the photonuclear cross-sections for $^{160}$Gd.
The arrows with separation energies indicate the thresholds for the 
$(\gamma,x$n$+y$p) reactions. Close to threshold a transmission factor of 
the neutron and the proton reduces the cross-section. The protons in addition
have a reduction by a Coulomb tunneling factor 
$\exp\left(-\frac{\pi (Z-1) e^2}{\hbar v}\right)$ with the velocity $v$ 
of the proton and the charge $Z$ of the nucleus, where Coulomb hindrance 
prevails up to the Coulomb energy $E_c = \frac{(Z-1)^2 e^2}{R}$ with 
the nuclear radius $R$. The exponential rise of the starting 
$(\gamma,x$n) reaction cross-sections is due to the increase in 
compound nucleus resonance level density.

If we could look with higher resolution into the photonuclear cross-sections,
we would observe individual resonances characterized by a width $\Gamma$.
The cross-section for a compound nucleus resonance of the ($\gamma$,x) 
reaction at the resonance energy $E_r$ is given by the Breit-Wigner 
formula \cite{segre77}:

\begin{equation}
   \sigma (E_{\gamma})= (\lambda_{\gamma}^2/4\pi)\cdot g \cdot
   \frac{\Gamma_{\gamma}\Gamma_d}{(E_{\gamma}-E_r)^2 +
   (\Gamma)^2/4}
\label{eq4}
\end{equation}


$g$ is a spin factor close to unity. 
$\lambda_{\gamma}=\hbar/(E_{\gamma}\cdot c)$ represents the
wavelength of the $\gamma$ rays with energy $E_{\gamma}$.
$\Gamma$ is the total width of the resonance with 
$\Gamma=\Gamma_{\gamma}+\Gamma_d+\Gamma_D$ and the decay widths 
$\Gamma_d$ to the desired product and $\Gamma_D$ to all other exit 
channels. $\Gamma_D$ may become important for less favored reactions
like ($\gamma$,p).

The width $\Gamma_{\gamma}$ has been studied systematically
as a function of $A$ at the neutron separation energy \cite{segre77} and
we obtain an average $\Gamma_{\gamma}\approx$ 100 meV for nuclei with $A=160$. 
Frequently the integrated cross-section, i.e. the product of cross-section 
times the width $\Gamma$ is given, which in our case is about 2 b$\cdot$eV.
 
\begin{figure}[t!]
\centerline{\includegraphics[width=.4\textwidth]{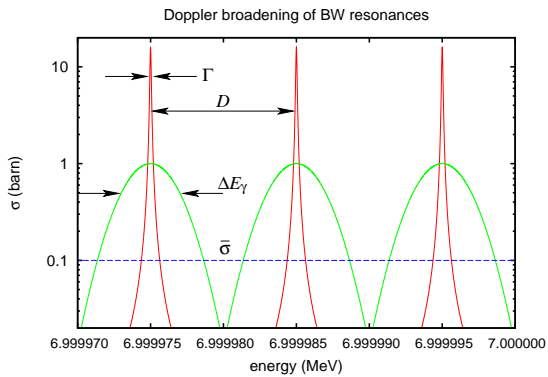}}
   \caption{ Breit-Wigner cross sections (red), Breit-Wigner cross section
             broadened by the Doppler effect (green) and the energy averaged
             cross sections ( dashed blue), showing the much larger individual
             resonance cross sections.}
   \label{fig5}
\end{figure}

The energy spacing of the compound nuclear resonances for a given spin 
and parity at the neutron binding energy for $A=160$ is about $D \approx 10$~eV
\cite{segre77}. It can be calculated from the back-shifted Fermi-gas formula
\cite{egidy05}. Thus with $\Gamma / D \approx$ 1\% probability we hit a 
resonance. In Fig.~\ref{fig5} we show these Breit-Wigner resonances.

For a given $\gamma$ beam energy of 7 MeV, a bandwidth $\Delta E \approx 7$~keV
will cover about 700 resonances. The width $\Gamma_{\gamma}$, where we 
have shown the average in Fig. \ref{fig4}, has a Porter-Thomas 
distribution \cite{weidenmueller09,richter10}

\begin{equation}
P(s)=\frac{1}{\sqrt{2\pi s}} \cdot \exp(-s/2).
\end{equation}

with $s=\Gamma_{\gamma}/<\Gamma_{\gamma}>$.
So most of the resonances have a very small $\gamma$ width and very few levels
show a much larger width. Thus from energy bin to energy bin we expect large
fluctuations of the average value within the bin and we can select an energy 
bin with a large cross section. The smaller the bandwidth of the $\gamma$-beam,
the larger these fluctuations become and one may select e.g. bins with 10 
times larger average cross section. Since the level spacings $D$ grow 
exponentially when reducing the mass number A at the same excitation energy, 
these fluctuations become more pronounced for lighter nuclei.

The Doppler broadening of a $\gamma$ transition at room temperature
$kT=1/40$ eV for a nucleus with mass number $A=160$ and  a $\gamma$ energy
$E_{\gamma}$= 7 MeV is

\begin{equation}
\Delta E_{\gamma}=E_{\gamma}\sqrt{(2kT)/m_u c^2 A}\approx \mbox{4
eV}
\end{equation}

Thus the line is broadened with respect to the natural linewidth 
by a factor of $\approx 40$. 

Here the spectral flux density $\Phi/(\Delta E)$ is an important quantity,
where we will reach values of $10^{13}\gamma/$(cm$^2 \cdot $s$ \cdot $eV) 
at ELI-NP, but up to $10^{17}\gamma/($cm$^2 \cdot $s$ \cdot $eV) in Compton 
back-scattering with energy recovery electron linacs.
The specific activity at saturation can then be calculated by multiplying 
the integrated cross-section $\sigma(E_r) \cdot \Gamma$ with the probability 
to hit a resonance and the spectral flux density:

\begin{equation}
\frac{A}{m} = \frac{N_A}{M} \left[\sigma(E_r) \cdot \Gamma\right]
\frac{\Gamma}{D} \cdot \left(\frac{\Phi}{\Delta E}\right)
\label{eq5}
\end{equation}


\section{Comparison of the Energy Loss in the Target between Photonuclear
and Ion-induced Reactions}

Gamma rays deposit their energy in quantized interactions with matter, such as
Compton scattering, pair creation, photo effect or photonuclear reactions.
For photon energies between 10 and 30 MeV, the total cross-section 
is dominated by Compton scattering and pair 
production in the nuclear field. The dependence
of the cross sections as a function of $\gamma$ energy and target element 
is given in Ref. \cite{Hub80}. The Compton cross-section rises proportionally 
to the number of electrons per atom, i.e. to $Z$, while the cross-section for 
pair production rises roughly with $Z^2$. At 15 MeV $\gamma$ energy we find 
for carbon a total cross section of 0.34 b/atom, while for lead this value 
amounts to 20 b/atom. Generally the cross section is rather flat in this 
energy region. While for carbon the Compton cross-section is twice the pair 
creation cross section, for lead the pair creation is five times more 
important than the Compton effect.

 
For 10 MeV $\gamma$ quanta the angle of the Compton scattered $\gamma$'s
is confined to about 10 degrees and the cross-section is strongly 
peaked in forward direction with an energy loss of less than 300 keV.
If we assume a typical total cross-section of 10 b/atom and a target
thickness of $10^6$ atomic layers, about 5\% of the $\gamma$ quanta will suffer
an energy loss by Compton scattering of 100 keV and about 5\%
will undergo pair creation at 10 MeV. However, 
in such thin targets of less than 0.1~g/cm$^2$ less
than $10^{-2}$ of the electrons are stopped and less than $5\cdot 10^{-5}$  
of the energy is deposited.

In contrast to gamma rays, charged particles deposit their energy continuously
while being slowed down in matter. The mean rate of energy loss 
is given in units of (MeV/(g/cm$^2$)) (stopping power) by the 
Bethe-Bloch equation:

\begin{equation}
   -\frac{d E}{d x}= K z^2\frac{Z}{A}\frac{1}{\beta^2}\left[\frac{1}{2}
  \ln \frac{2 m_e c^2 \beta^2 \gamma^2 T_{\rm max}}{I^2}-\beta^2 -
\frac{\delta}{2}\right]
\end{equation}

with K=$4\pi N_A r_e^2 m_ec^2$=0.307 MeV cm$^2$. Here $T_{\rm max}$ is the
maximum kinetic energy, which can be transfered in a single collision,
$Z$ the atomic number of the target, $A$ the atomic number of the target,
$ ze$ the charge of the ion and $\delta$ a density effect correction to the
ionization energy loss. For 10 MeV protons we obtain 30 MeV/(g/cm$^2$).
Therefore with the density of iron of 7.9 g/cm$^3$ we have 
an energy loss of 10 MeV in 0.26 mm.
In these $2.6\cdot 10^6$ atomic layers at \AA ~distances,
about $10^{-8}$ reactions occur per atomic layer or 2.6\% in total. 
Thus we deposit per produced new radioactive
nucleus about 400 MeV. The energy deposition is about a factor 
of $10^5$ larger for protons compared to $\gamma$'s for the same number 
of produced nuclei. 

The typical intensity of proton beams used for isotope production is 
of the order of 100 $\mu$A/cm$^2$, corresponding to $6\cdot 10^{14}$/(cm$^2$s).
On the other hand, the target should withstand a higher $\gamma$ flux density 
of 10$^{19}$/(cm$^2$s). However, for Bremsstrahlung beams one has a strong 
rise of the $\gamma$ ray spectrum to low energies with increased energy 
deposition at lower energies, making it worse compared to proton activation.

\section{Specific Radioisotopes produced in Photonuclear Reactions}

We now discuss in detail the different $\gamma$-induced reactions and specific
radioisotopes that can be produced by such reactions.

In tables \ref{isomer-table} and \ref{production-table} we show estimates 
of the achievable specific activities for thin targets for a $\gamma$ flux 
of $10^{14}$ per s, corresponding to a flux density of $10^{18} \, 
\gamma/$(cm$^2$ s). With a bandwidth of $10^{-3}$ this results
at 10~MeV in a spectral flux density of $10^{14} \, \gamma/$(cm$^2$ s eV).
These values are inbetween the characteristics of presently built facilties 
(e.g. MEGa-RAY) and future improved facilities (100 mA ERL). We compare these 
to thin target yields obtained by thermal neutron capture in a typical 
flux density of $10^{14}$ n/(cm$^2$ s) in high flux reactors. Note that 
alike for the potential beam parameters of $\gamma$ beam facilities, there 
is also a wide range for the flux density really available at the irradiation 
positions of high flux reactors. Some positions provide flux densities of 
several $10^{12}$ to $10^{13}$ n/(cm$^2$ s), while few special reactors 
have positions that even exceed $10^{15}$ n/(cm$^2$ s), namely SM3 in 
Dimitrovgrad \cite{Kar97}, HFIR in Oak Ridge \cite{Kna05}
and ILL's high flux reactor in Grenoble.

Since hitherto no $\gamma$ beams with sufficiently small bandwidth were 
available to exploit resonant excitation, there are obviously no such 
measured cross-sections. Hence, we can only estimate a lower 
bound using the averaged cross-sections measured at Bremsstrahlung 
facilities \cite{data,Neu91,Car91,Car91a,Car93}. For cases where no measured cross-sections 
are available, we interpolate experimental cross-sections of the same 
reaction channel on nearby elements, taking into account the energy above 
the reaction threshold.

Even when using these conservative assumptions, the estimated specific 
activities are very promising for specific isotopes.

The total activity in a nuclear reactor can be relatively high since 
thick (several cm) and large (several cm$^2$) targets can be used if 
the cross-sections are not too high (leading to self absorption and local 
flux depression). Multiple irradiation positions allow producing various 
radioisotopes with activities of many TBq.

For the $\gamma$ beam we estimate the total activities by integrating to 
one interaction length, i.e. where the initial $\gamma$-beam intensity 
has dropped to $1/e = 37$~\% of its intensity. Higher total activities 
can be achieved with thicker targets at the expense of lower specific 
activity and vice versa. The total interaction cross-section is usually 
dominated by the atomic processes of Compton effect and pair creation. 
We consider conservatively any interacted $\gamma$ ray as lost. 
In reality, part of the Compton scattering goes forward under small angles 
and the $\gamma$ rays that have lost little energy can still induce 
photonuclear reactions. The usable target thickness ranges from 
20~g/cm$^2$ for heavy elements to 40~g/cm$^2$ for light elements, i.e.
in total only few mg target material are exposed to the small area 
of the $\gamma$ beam. With non-resonant reactions of the order of 
0.1 TBq activity can be produced per day, corresponding to tens 
(for $\beta^-$ therapy isotopes) to thousands (for imaging isotopes 
and therapy with alpha emitters) of patient doses.

We consider metallic targets, although in praxi for certain elements 
rather chemical compounds would be used as targets. Usually oxides or
other compounds with light elements are used. These light elements have 
a relatively low cross-section for gamma rays, hence the specific 
activity achieved with compound targets is not much lower compared to 
elemental targets.

The exact target geometry does not affect our estimates. In principle 
a single compact target or a stack of thin target foils could be used 
and would provide similar production rates. In practice the latter 
solution can stand far higher beam intensities. The foils could be 
radiation-cooled in vacuum or helium-cooled, since helium has a low $Z$ 
and correspondingly low cross section for interaction with gamma rays. 
Due to the low divergence of the $\gamma$ beam, the individual target 
foils can be spaced wide apart, thus reducing the view factors between 
the foils to minimize mutual heating by radiation absorption.
For sufficiently thin foils most of the forward-directed Compton and 
pair electrons and positrons can leave the foil. Spacing the foils further 
apart reduces the energy deposition from electrons of the previous foil, 
which deposit their energy laterally (e.g. in a water-cooled target chamber)
spread over a wide area. The trajectories of the electrons and positrons 
could even be forced outward by applying transversal magnetic fields.
Also a stack of target foils with thin water-cooling channels in between 
can be considered since hydrogen and oxygen have much lower interaction 
cross-sections with gamma rays.

Alternatively to thin foils also a thin wire (e.g. 0.1 mm diameter) 
or several consecutive wires could be placed along the $\gamma$ beam 
direction. Here most electrons and positrons that are emitted under angles 
different from 0 degree will rapidly leave the target and not contribute much 
to its heating. Even those that are initially emitted forward will rapidly 
change direction by scattering and then leave the wire. In particular for 
less intense $\gamma$ beams such a solution may 
be simpler to realize than a multi-foil stack.

All these heat dissipation techniques rely on the small area, small 
divergence and small bandwidth of a $\gamma$ beam. They could not be 
applied for Bremsstrahlung spectra. Thus, the extremely high
flux densities of $\gamma$ beams can really be utilized without 
being seriously limited by the required heat dissipation from the 
targets as is frequently the case for charged-particle induced reactions
or intense Bremsstrahlung spectra.

Instead of producing a single product isotope at a time, the target 
stack could also consist of different targets for simultaneous production 
of different isotopes. This is possible when the different reactions 
require similar $\gamma$ energies. It may be particularly efficient 
when at least one of the reactions is characterized by prominent 
resonances, reducing the interaction length for resonant $\gamma$ rays. 
The ``unused'' $\gamma$ rays within the bandwidth of the $\gamma$ beam 
may then be used downstream for other reactions that are not resonant or have
resonances at different energies.

\subsection{\boldmath \bf Isomers of stable isotopes via $(\gamma,\gamma ')$ 
reactions \unboldmath}

\begin{table*}[ht]
\caption{Longer-lived nuclear isomers produced in ($\gamma,\gamma '$)
reactions. The relative population of the respective isomer in thermal 
neutron capture on $A$-1 target isotopes is given as $I_{is}/I_{gs}$ 
where known experimentally. Experimental integrated cross sections for 
population of the isomer by ($\gamma,\gamma')$ reactions at 4 MeV and 
6 MeV were taken from \cite{Car91,Car91a}. The fraction of the maximum 
specific activity produced in $(\gamma,\gamma')$ reactions $R_{\gamma}$, 
is put in relation to that obtained with (n,$\gamma)$ reactions 
$R_{({\rm n},\gamma)}$ at a thermal neutron flux of $10^{14}$ n./(cm$^2$ s) 
in the last column.}
\label{isomer-table}
\begin{center}
\setlength{\tabcolsep}{1mm}
\begin{tabular}{c|ccc|c|cc|cc|ccc|c|c} \hline
\hline
Iso-	&	\multicolumn{3}{c|}{Isomer} & 	$I_{is}/I_{gs}$	&	\multicolumn{2}{c|}{Ground state}	&	\multicolumn{2}{c|}{$\sigma \cdot \Gamma$}	&	Spec. act.	&	Activity & $R_\gamma$	&	Spec. act.	&	$R_\gamma / R_{({\rm n},\gamma)}$	\\
tope	&	Exc.	&	Spin \&	&	$T_{1/2}$	&	& Spin \&	&	Nat.	&	at	&	at	&	$\gamma$ beam	&	per day	&	fraction&	high flux	&		\\
	&	energy	&	parity	&		&		&	parity	&	abun.	&	4 MeV	&	6 MeV	&	& & of max.	&	reactor	&		\\
	&	keV	&		&	d	&		&		&	\%	&	eV$\cdot$b	&	eV$\cdot$b	&	GBq/mg	&	GBq	&		&	GBq/mg	&		\\
\hline
$^{87}$Sr	&	389	&	1/2$^-$	&	0.12	&	3.5	&	9/2$^+$	&	7	&	3.9	&	8.7	&	10	&	110	&	2$\cdot$10$^{-5}$	&	0.57	&	18	\\
$^{115}$In	&	336	&	1/2$^-$	&	0.19	&		&	9/2$^+$	&	95.7	&	18	&	67	&	58	&	603	&	3$\cdot$10$^{-4}$	&		&		\\
$^{117}$Sn 	&	315	&	 11/2$^-$ 	&	13.8	&	0.04	&	1/2$^+$ 	&	7.68	&	3.2	&	8.8	&	7.5	&	0.6	&	0.0025	&	0.003	&	2400	\\
$^{119}$Sn 	&	90	&	 11/2$^-$ 	&	293	&	0.02	&	1/2$^+$ 	&	8.59	&		&		&		&		&		&	0.002	&		\\
$^{123}$Te 	&	248	&	 11/2$^-$ 	&	119	&	0.13	&	1/2$^+$ 	&	0.89	&	42	&	68	&	55	&	0.6	&	0.17	&	0.2	&	280	\\
$^{125}$Te 	&	145	&	 11/2$^-$ 	&	57.4	&	0.17	&	1/2$^+$ 	&	7.07	&	70	&		&		&		&	0.08	&	0.48	&		\\
$^{129}$Xe 	&	236	&	 11/2$^-$ 	&	8.9	&	0.10	&	1/2$^+$ 	&	26.4	&	&		&		&		&		&	0.2	&		\\
$^{131}$Xe 	&	164	&	 11/2$^-$ 	&	11.8	&	0.10	&	3/2$^+$ 	&	21.2	&	&		&		&		&		&	0.2	&		\\
$^{135}$Ba 	&	268	&	 11/2$^-$ 	&	1.2	&	0.08	&	3/2$^+$ 	&	6.59	&	13	&	60	&	44	&	33	&	1.5$\cdot$10$^{-3}$	&	0.045	&	1000	\\
$^{176}$Lu	&	123	&	1$^-$	&	0.15	&	2.0	&	7$^-$	&	2.59	&		140	&	350	&	2.0	&	1800	&	1.2$\cdot$10$^{-3}$	&	5.5	&	36	\\
$^{180}$Hf 	&	1141	&	 8$^-$    	&	0.23	&	0.01	&	0$^+$   	&	35.1	&	&		&		&		&		&	0.14	&		\\
$^{193}$Ir 	&	80	&	 11/2$^-$ 	&	10.5	&		&	3/2$^+$ 	&	62.7	&		&		&		&		&	10$^{-5}$	&		&		\\
$^{195}$Pt 	&	259	&	 13/2$^+$ 	&	4.02	&	0.09	&	1/2$^-$ 	&	33.8	&	30	&	140	&	72	&	17	&	0.012	&	0.019	&	3800	\\
\hline
\end{tabular}
\end{center}
\end{table*}

Longer-lived nuclear isomers that decay by emission of gamma rays and/or 
conversion electrons to the respective ground state are of interest 
for various applications in nuclear medicine if they can be produced 
with high specific activity. Table \ref{isomer-table} shows a selection 
of such isomers. The table shows in addition also other isomers that 
have at present no application in nuclear medicine, but are of interest 
e.g. for M\"o\ss{}bauer spectroscopy or even gamma-ray lasers.

Most usual production methods, e.g. via (n,$\gamma$) reactions, 
result in relatively low specific activity since the dominant part 
of the production proceeds directly to the nuclear ground state 
that has a nuclear spin closer to that of the $A$-1 target isotope. 
However, the fact that all these isomers are actually populated via 
thermal neutron capture reactions on low-spin $A$-1 target isotopes 
proofs that pathways populating the high-spin isomers from higher-lying, 
low-spin compound nucleus resonances levels of lower spins must exist.
In Ref. \cite{ledoux06} the population of high-spin isomers relative 
to the ground state was studied for resonances in (n,$\gamma$) reactions. 
An energy dependence of the isomeric ratio was observed. 
One may expect that this energy dependence would become even more 
pronounced if the reactions were excited with a primary beam of smaller 
bandwidth that populates more selectively states which decay mainly
to the isomeric level of interest.

Also photoexcitation ($\gamma,\gamma'$) experiments with Brems\-strahlung 
beams were performed on a series of stable targets and showed strong 
population of isomeric levels \cite{Neu91,Car91,Car93}. 
The observed energy dependence of the isomer activation yields indicates 
that few gateway states are responsible for populating efficiently the isomers.

It was recently demonstrated that Coulomb excitation to states communicating 
by gamma transitions with different isomeric states of a nucleus can be 
used to modify the initial isomeric composition \cite{Ste07}. We propose 
to use a similar method, but using gamma rays with small bandwidth directed 
onto the stable target nuclei. Compared to Coulomb excitation or 
Bremsstrahlung spectra, the excitation is far more selective 
and a much higher conversion rate can be achieved due to the high 
gamma ray flux density.

Moreover, photoexcitation with small bandwidth $\gamma$ rays allows moreover 
the selective excitation of individual levels or groups of levels 
that decay preferentially to the nuclear isomer, thus enhancing the 
specific activity of the isomer. Only in few cases the energies of 
such (groups of) levels are already known. Note that relatively low 
gamma ray energies may be sufficient for such a pumping to isomeric states.
In $^{125}$Te a $7/2^+$ state at only 402 keV excitation energy can 
serve as gateway state for pumping from the $1/2^+$ ground
state to the $11/2^-$ isomer at 145 keV \cite{NuDat}. 

We will estimate the achievable specific activity at the example 
of $^{115}$In for which the required transition energies, branching ratios 
and transition strengths are already experimentally known, even if this 
isomer has presently no application in nuclear medicine.
The $9/2^+$ ground state of $^{115}$In can be excited by an $E2$ 
transition from the ground state to the $5/2^+$ level at 1078.2~keV 
that decays with 16\% cumulative branching ratio to the $1/2^-$ isomer 
at 336 keV \cite{NuDat}. The gateway level at 1078~keV has a half-life 
of 0.99~ps, corresponding to a total width $\Gamma$ of 0.67~meV. 
The fractional width to the ground state $\Gamma_\gamma$ is 0.55~meV. 
while the decay to the isomer has a width of $\Gamma_d = 0.10$~meV. 
Hence, with the formula of equation \ref{eq4} we obtain at the resonance energy
$E_r = 1078$~keV a peak cross-section of $\sigma (1078 \mbox{ keV}) = 3400$~b.
The product of peak cross-section times width is 2.3 eV$\cdot$b.
With a $\gamma$ beam flux density per eV of $10^{16}\gamma/($cm$^2 
\cdot $s$ \cdot $eV) we obtain $^{193m}$Ir with a specific activity 
of 20~GBq/mg at saturation, i.e. $R_\gamma = 10^{-4}$.

In many nuclei only a small fraction of excited levels and transitions 
between them are experimentally known, in particular for higher 
excitation energies. Also present theoretical models do not allow 
to predict the exact position and decay modes of higher-excited levels.
Only in particular cases the population of high-spin isomers can be 
explained with individual intermediate levels from experiment and theory, 
see e.g. \cite{Neu91,Car93,Bon99}.

Experimental data on isomer population by ($\gamma,\gamma'$) reactions 
had so far been obtained with Bremsstrahlung spectra of large bandwidth. 
The integrated cross-sections at $\gamma$ energies of 4 and 6~MeV, 
respectively, are of the order of 10 to 100 b$\cdot$eV.

Many potential gateway states that could serve for pumping nuclei from 
their ground state to isomeric levels are expected to exist, but they 
still need to be identified by dedicated high resolution measurements from 
excitation energies of few hundred keV up to close to the particle 
separation energy. Thes measurements have to be performed with the new 
$\gamma$ beams for each of the isotope for several thousand energy windows, in
order to determine the best excitation deexcitation path to the isomer. 
Presently existing $\gamma$-ray beam facilities do not provide 
sufficiently monochromatic $\gamma$-ray beams to search for suitable 
resonance regions. A systematic investigation will require Compton 
backscattering facilities such as MEGa-Ray or ELI-NP.

Selecting $\gamma$ ray energies providing strong pumping to 
the isomeric state will improve the achievable specific activity 
correspondingly. Even multiple excitations of the path to the 
isomer are possible. Due to the missing energy match, no significant 
back-pumping from the isomer to the ground state will occur.

Note that for populating isomeric levels in ($\gamma,\gamma '$) reactions 
they do not necessarily need to be excited close to the neutron binding 
energy. To prevent competing photonuclear reactions, it is actually more
favorable to select the lowest excited levels that decay populate with 
sufficiently high yield the isomeric level. 


Two examples of long-lived isomers with important medical applications 
are discussed in the following:

\begin{enumerate}

\item $^{195m}$Pt:\\
Platinum compounds such as cisplatin or carboplatin are known to be 
cytotoxic and are frequently used for chemotherapy of tumors. Labeling 
these compounds with platinum radiotracers allows for in-vivo 
pharmacokinetic studies and tumor imaging, e.g. to monitor the 
patient-specific uptake and optimize the dosing individually \cite{Dow00}. 
Failure to demonstrate the tumor uptake of the chemotherapy agent by 
nuclear imaging helps to exclude those ``non-responding'' patients 
from unnecessary chemotherapy treatment. $^{195m}$Pt has 4 days half-life 
and emits a 99~keV gamma ray that can be used for imaging by SPECT or 
gamma cameras. $^{195m}$Pt emits also low-energy conversion and Auger 
electrons. Hence, when used in higher activities, it could 
be suitable for a combined chemo- and radionuclide therapy.
Unfortunately $^{195m}$Pt is destroyed by (n,$\gamma$) reactions 
with a very high cross section of 13000~barn. Therefore the specific 
activity achievable by neutron capture on $^{194}$Pt is seriously limited. 
Even at the HFIR reactor in Oak Ridge only 0.04 GBq/mg are obtained 
\cite{Kna05} and too little activity is presently available for clinical 
trials \cite{Riv05}. By ($\gamma,\gamma'$) reactions we expect to obtain 
much higher specific activities, namely about 70 GBq/mg! About 20 GBq could 
be produced per day, sufficient for several hundred patient-specific uptake 
measurements or to launch first clinical trials for radionuclide therapy 
with $^{195m}$Pt. Moreover, even if natural platinum or platinum compounds 
are irradiated, the radionuclidic purity of the product 
will be excellent since no other long-lived radioisotopes can be 
produced by activation with few MeV gamma rays.

\item $^{117m}$Sn:\\
Also $^{117m}$Sn emits low-energy conversion and Auger electrons, making 
it promising for radionuclide therapy. In addition it emits a 159 keV 
gamma ray for imaging. It has been shown that $^{117m}$Sn can be used 
for pain palliation in bone metastases of various cancers.
Due to its soft electron energy spectrum it has less side effects on 
the bone marrow than other radioisotopes with more penetrating radiation 
\cite{Bis00}. Unfortunately the high-spin isomer $^{117m}$Sn is poorly 
produced in thermal neutron capture on zero-spin $^{116}$Sn.
With inelastic neutron scattering $^{117}$Sn(n$_{\rm fast}$,n'$\gamma$) 
specific activities of 0.2 to 0.4 GBq/mg are obtained at high flux 
reactors \cite{Kna05,Pon09}, but too little activity is presently 
available \cite{Riv05}. Production via ($\gamma,\gamma'$) reactions 
with 6~MeV $\gamma$ beams allows boosting the specific activity at least 
to 7~GBq/mg. About 0.6~GBq $^{117m}$Sn are produced per day, sufficient 
to start clinical trials.
\end{enumerate}

These two isomers appear at present most interesting for nuclear 
medicine applications. The specific activity and total production 
per day could be significantly improved with still to be found better 
gateway states. A detailed search for suitable gateway states at 
an upcoming $\gamma$-beam facility with small bandwidth is urgently needed.

\subsection{\boldmath \bf Radioisotopes via the ($\gamma$,n) reaction 
 \unboldmath}

\begin{table*}[ht]
\caption{Estimated production rates of radioisotopes produced in ($\gamma$,n), 
($\gamma$,p) or ($\gamma$,2n) reactions. 
Experimental cross sections were taken from \cite{data}, estimated 
cross sections are marked in italics.
The fraction of the maximum specific activity produced in $(\gamma$,x) 
reactions $R_{\gamma}$, is put in relation
to that obtained with (n,$\gamma)$ reactions $R_{({\rm n},\gamma)}$ at 
a thermal neutron flux of $10^{14}$ n/(cm$^2$ s) 
in the last column. $^*$: For comparison we show the values for $^{99}$Mo 
produced by $^{98}$Mo(n,$\gamma$). However,
usually $^{99}$Mo is produced by fission with much better specific activity.}
\label{production-table}
\begin{center}
\setlength{\tabcolsep}{1mm}
\begin{tabular}{c|ccccc|ccc|c|c} \hline
\hline																		
Product	&	$T_{1/2}$	&	Target	&	Reaction	&	$E_\gamma$	&	$\sigma$	&	Spec. act.	&	Activity	&	$R_\gamma$	&	Spec. act.	&	$R_\gamma / R_{({\rm n},\gamma)}$	\\
isotope	&		&	isotope	&		&		&		&	$\gamma$ beam	&	per day	&	fraction	&	high flux reactor	&		\\
	&	d	&		&		&	MeV	&	b	&	GBq/mg	&	GBq	&	of max.	&	GBq/mg	&		\\
\hline																		
$^{47}$Ca	&	4.5	&	$^{48}$Ca	&	($\gamma$,n)	&	19	&	0.09	&	1100	&	400	&	0.05	&	0.9	&	1200	\\
$^{64}$Cu	&	0.5	&	$^{65}$Cu	&	($\gamma$,n)	&	17	&	0.09	&	830	&	1150	&	0.006	&	4	&	200	\\
$^{99}$Mo	&	2.8	&	$^{100}$Mo	&	($\gamma$,n)	&	14	&	0.16	&	960	&	350	&	0.06	&	0.08$^*$	&	12000	\\
$^{103}$Pd	&	17	&	$^{104}$Pd	&	($\gamma$,n)	&	17	&	{\it 0.05}	&	290	&	16	&	0.1	&	1.8	&	160	\\
$^{165}$Er	&	0.4	&	$^{166}$Er	&	($\gamma$,n)	&	13	&	0.3	&	1100	&	1100	&	0.016	&	4.7	&	230	\\
$^{169}$Er	&	6.9	&	$^{170}$Er	&	($\gamma$,n)	&	12	&	{\it 0.3}	&	$\approx 800$	&	130	&	$\approx 0.2$	&	0.8	&	1000	\\
$^{186}$Re	&	3.7	&	$^{187}$Re	&	($\gamma$,n)	&	15	&	0.6	&	$\approx 1400$	&	320	&	$\approx 0.2$	&	35	&	40	\\
$^{225}$Ra	&	14.8	&	$^{226}$Ra	&	($\gamma$,n)	&	12	&	{\it 0.2}	&	$\approx 300$	&	30	&	$\approx 0.2$	&		&		\\
\hline
$^{47}$Sc	&	3.4	&	$^{48}$Ti	&	($\gamma$,p)	&	19	&	{\it 0.02}	&	250	&	100	&	0.009	&		&		\\
$^{67}$Cu	&	2.6	&	$^{68}$Zn	&	($\gamma$,p)	&	19	&	{\it 0.03}	&	260	&	115	&	0.01	&		&		\\
\hline
$^{44}$Ti	&	60 y	&	$^{46}$Ti	&	($\gamma$,2n)	&	27	&	{\it 0.01}	&	$\approx 0.5$	&	0.008	&	$\approx 0.1$	&		&		\\
$^{224}$Ra	&	3.7	&	$^{226}$Ra	&	($\gamma$,2n)	&	16	&	{\it 0.1}	&	$\approx 50$	&	10	&	$\approx 0.01$	&		&		\\
\hline																		
\end{tabular}
\end{center}
\end{table*}

When being excited well beyond the neutron binding energy a nucleus loses 
readily a neutron. Competing reactions such as deexcitation by gamma 
ray emission are far less probable.

\begin{enumerate}
\item $^{99}$Mo/$^{99m}$Tc:\\
The presently most used radioisotope for nuclear me\-di\-cine studies 
is $^{99m}$Tc. Its 140~keV $\gamma$ ray is ideal for SPECT imaging. With 
a relatively short half-life of 6 h and the quasi-absence of beta particles 
the radiation dose to the patient is low. $^{99m}$Tc is conveniently eluted 
in non-carrier-added qua\-li\-ty from simple and reliable 
$^{99}$Mo ($T_{1/2} = 66$~h) generators that can be used 
for about one week. Various technetium compounds have been developed 
for a multitude of nuclear medicine applications \cite{schiepers06}.
The combination of these advantages explains why $^{99m}$Tc is used 
in about 80\% of all nuclear medicine studies. 
28 millions applications employing $^{99m}$Tc are performed per year. 
Every week more than 80~kCi (3000 TBq) of $^{99}$Mo have to be produced 
to load the $^{99}$Mo/$^{\rm 99m}$Tc generators that are shipped to 
the $^{99m}$Tc users (mainly hospitals) or to radiopharmacies. 
Until recently five nuclear reactors were used to produce about 95\,\% of the 
world needs of $^{99}$Mo by neutron-induced fission of highly enriched 
$^{235}$U targets. Recently the two reactors that used to produce the 
majority of the $^{99}$Mo supply had extended shutdowns,
leading to a serious $^{99}$Mo/$^{\rm 99m}$Tc supply crisis \cite{Ral09,Lew09}.
This triggered large interest in alternative $^{99}$Mo/$^{\rm 99m}$Tc 
production paths. One alternative production path uses the 
$^{100}$Mo($\gamma$,n) reaction. Usual Bremsstrahlung facilities 
produce $^{99}$Mo with limited specific activity that makes it more 
difficult to use the established generator technology (acid alumina columns). 
With $\gamma$ beams of high flux density, $^{99}$Mo could be produced 
with much higher specific activity, allowing direct use of existing 
generator technology. A facility providing $10^{15}\gamma$/s could 
produce via $^{100}$Mo($\gamma$,n) reactions several TBq per week. 
Thus, many such facilities would be required to assure the 
worldwide $^{99}$Mo supply. 

\noindent This example demonstrates that the new production method 
by $\gamma$ beams is not intended to compete with large-scale production 
of established isotopes. The advantage of $\gamma$ beams for radioisotope 
production lies clearly in the very high specific activity that can be 
achieved for radioisotopes or isomers that are very promising for nuclear 
medicine, but that are presently not available in the required quality or 
quantity. Examples of such isotopes and new clinical applications that 
may become available with such radioisotopes produced by $\gamma$ beams 
will be discussed in the following.

\item $^{225}$Ra/$^{225}$Ac:\\ 
Alpha emitters are very promising for therapeutic applications,
since the emitted alphas deposit their energy very locally
(typical range of one to few cancer cell diameters) with high linear energy
transfer (LET) and, hence, high probability for irreparable double strand
breaks. An alpha emitter coupled to a cancer cell specific bioconjugate can
be used for targeted alpha therapy to treat disseminated cancer types
(leukemia), micro-metastases of various cancers or to destroy chemo- and
radiation-resistant cancer cells (e.g. glioblastoma). One pro\-mising
alpha emitter is $^{225}$Ac ($T_{1/2}=10$~days) that decays by a series 
of four alpha decays and two beta decays to $^{209}$Bi, see 
table \ref{Np-chain}. $^{225}$Ac (valence III) can either be used directly 
for targeted alpha therapy, or as generator for $^{213}$Bi that 
is used for targeted alpha therapy. Today $^{225}$Ac is produced by 
decay of $^{229}$Th$\rightarrow ^{225}$Ra$\rightarrow ^{225}$Ac and 
chemical separation. Unfortunately only small quantities 
(about 1 Ci=37 GBq per year) are available \cite{Apo01}, 
which is far too little for a large scale application. Alternatively, 
$^{226}$Ra can be converted by (p,2n) reactions to $^{225}$Ac \cite{Apo05} 
or by ($\gamma$,n) reactions to $^{225}$Ra that decays to $^{225}$Ac and is 
subsequently chemically separated from the $^{226}$Ra target \cite{maslov06}. 
The radioactive $^{226}$Ra targets are difficult to handle, when the 
activity of the target gets important. Therefore a $\gamma$ beam with 
high flux density is essential to minimize the target size 
and target activity while maximizing the product activity.
Thus about 200 GBq $^{225}$Ac could be produced per week, enough to treat
hundreds of patients.

\begin{table}[t]
\caption{Radioisotopes of the neptunium chain produced in 
the $^{226}$Ra($\gamma$,n)$^{225}$Ra reaction and subsequent decays.}
\label{Np-chain}
\begin{center}
\begin{tabular}{cccc} \hline
Isotope   & $\alpha$ energy & Mean $\beta$ energy &$T_{1/2}$\\ \hline
& (MeV) & (MeV) & \\
$^{225}$Ra&                      & 0.1           & 15 d    \\
$^{225}$Ac&  5.8                 &                     & 10 d    \\
$^{221}$Fr&  6.3                 &           & 4.9 min \\
$^{217}$At&  7.1                 &           & 32 ms    \\ 
$^{213}$Bi&  5.8 (2\%)                    & 0.4 (98\%)      & 46 min   \\
$^{213}$Po&  8.4                 &           & 4.2 $\mu$s\\
$^{209}$Pb&                      & 0.2       & 3.25 h     \\  \hline
\end{tabular}
\end{center}
\end{table}

\item $^{169}$Er:\\ 
$^{169}$Er decays with 9.4 days half-life by low-energy beta emission 
(100~keV average beta energy). These betas have a range of 100 to 200 
$\mu$m in biological tissue, corresponding to few cell diameters.
The short beta range makes this isotope very interesting for targeted 
radiotherapy \cite{Uus06}. However, due to the low 
$^{168}$Er(n$_{\rm th},\gamma$) cross section it cannot
be produced with high specific activity\footnote{Today $^{169}$Er with low 
specific activity is actually used for radiation synovectomy to treat 
inflamed small joints, e.g. in fingers \cite{Kar10}.} by neutron capture. 
Using intense $\gamma$ beams one can reach significantly higher 
specific activities via $^{170}$Er($\gamma$,n) reactions.

\item $^{165}$Er:\\ 
$^{165}$Er is one example for an isotope that decays mainly by low-energy 
Auger electrons. Their range is shorter than one cell 
diameter. Hence, these Auger emitters have to enter the cell and approach
the cell's nucleus to damage the DNA and destroy a cell. Coupled to a 
bioconjugate that is selectively internalized into cancer cells it can 
enhance the ratio for dose equivalent delivered to the tumor cell with 
respect to normal cells. This should result in an improved tumor treatment 
with less side effects. R\&D to identify suitable bioconjugates is under 
way \cite{Buc06}.

\item $^{47}$Sc:\\ 
$^{47}$Sc is a promising low-energy beta emitter for targeted radiotherapy. 
Scandium is the lightest rare earth element. Most established labeling 
procedures for valence III metals (Y, Lu,\dots) can be applied directly 
for Sc. Its 159~keV gamma line allows imaging of $^{47}$Sc distribution 
by SPECT or gamma cameras. Alternatively the $\beta^+$ emitting scandium 
isotope $^{44}$Sc can be used for PET imaging as a ``matched pair''. 
Carrier-free $^{47}$Sc can be produced by $^{50}$Ti(p,$\alpha$) or 
$^{47}$Ti(n$_{\mbox{fast}}$,p) reactions followed by chemical separation. 
The alternative production via 
$^{46}$Ca(n,$\gamma$)$^{47}$Ca$\rightarrow^{47}$Sc 
is uneconomic due to the extremely low natural abundance of $^{46}$Ca.
At present too little $^{47}$Sc is available \cite{Riv05}, but with 
intense $\gamma$ beams the production via 
$^{48}$Ca($\gamma$,n)$^{47}$Ca $\rightarrow~^{47}$Sc becomes competitive.
After grow-in of $^{47}$Sc it is chemically separated from the irradiated 
calcium, which, after sufficient decay of $^{47}$Ca, can be reused to form 
a new irradiation target.

\item $^{64}$Cu:\\
$^{64}$Cu is a relatively long-lived $\beta^+$ emitter ($T_{1/2}=12.7$~h) 
with various applications in nuclear medicine \cite{And09}. $^{64}$Cu-ATSM 
is a way to measure hypoxia of tumors. Hypoxia is an important effect 
influencing the resistance of tumor cells against chemo- or radiation therapy. 
$^{64}$Cu can also act itself as therapeutic isotope due to its emission 
of $\beta^-$ (191~keV mean energy) and low energy Auger electrons.
Today $^{64}$Cu is mainly produced with small cyclotrons by the 
$^{64}$Ni(p,n) reactions. Alternative production by $^{65}$Cu($\gamma$,n) 
does not require the rare and expensive $^{64}$Ni targets and saves the 
chemical separation step.

\item $^{186}$Re:\\
$^{186}$Re is a radioisotope suitable for bone pain palliation, 
radiosynovectomy and targeted radionuclide therapy.
Rhenium is chemically very similar to its homologue technetium, thus 
known compounds that have been developed for imaging 
with $^{99m}$Tc can also be labeled with $^{186}$Re and used for therapy. 
$^{186}$Re is currently either produced by neutron capture on $^{185}$Re, 
resulting in limited specific activity, or by $^{186}$W(p,n) reactions 
followed by chemical Re/W separation. The latter guarantees excellent 
specific activity at the expense of much reduced production
rates and a required chemical separation. Production by $^{187}$Re($\gamma$,n) 
would allow producing larger amounts (2 TBq per week) of $^{186}$Re with 
high specific activity. Enriched $^{187}$Re targets should be used to 
minimize contamination of the product with long-lived $^{184,184m}$Re 
by $^{185}$Re($\gamma$,n) reactions.
 
\item ``Slightly neutron-deficient radioisotopes'':\\
``Slightly neutron-deficient isotopes'' are decaying by electron capture 
  with emission of X-rays and low-energy Auger electrons, partially also
  gamma rays and conversion electrons. The absence of beta emission and the
  presence of low-energy X-rays or electrons is of advantage for a variety of
  applications such as calibration sources, radionuclide therapy
  applications after internalization into cells, etc. All these isotopes can
  be produced by neutron capture on the stable ($A$-1) neighboring
  isotope. However, as shown in Tab. \ref{slightly-n-def},
  the latter is generally very rare in nature (since only
  produced by unusual astrophysical processes like the p-process) and
  correspondingly costly when produced as isotopically enriched target
  material. Using instead ($\gamma$,n) reactions to populate the same isotopes
  allows using the much more abundant, and hence cheaper, ($A$+1) neighboring
  isotope as target. An example is $^{103}$Pd, a low-energy electron emitter. 
It can be used for targeted radionuclide therapy (coupled to a suitable 
bioconjugate) or for brachytherapy applications, where sources (``seeds'') 
are inserted into a cancer (e.g. breast cancer \cite{Bal10,Jan07}) for
localized irradiation. However, the target $^{102}$Pd for production by 
neutron capture is rare and expensive. Production via $^{104}$Pd($\gamma$,n) 
is more economic, if sufficiently intense $\gamma$ beams become available.
The same applies to other radioisotopes or isomers shown in 
table \ref{slightly-n-def} that have applications in nuclear medicine or 
other fields.
\end{enumerate}

\begin{table}[t!]
\caption{Comparison of target isotopes required for (n,$\gamma$) 
and $(\gamma$,n) reactions respectively leading to the
same radioisotopes or isomers. These have applications in nuclear 
medicine or other fields. The last column shows the ratio of the natural 
abundances of the $A$+1 and $A$-1 targets.}
\label{slightly-n-def}
\bigskip
\begin{center}
\begin{tabular}{cccccccc} \hline
Product   & $T_{1/2}$ & (n,$\gamma$) & nat. & ($\gamma$,n) & nat.& [$A$+1] \\
isotope   &           & target       & abun.  &  target      & abun. &    /         \\
          &  (d)   &              & (\%)    &              & (\%)   & [$A$-1]    \\
\hline
$^{47}$Ca & 4.5       & $^{46}$Ca    & 0.004   &  $^{48}$Ca    & 0.187  & 47\\
$^{51}$Cr & 27.7      & $^{50}$Cr    & 4.3     &  $^{52}$Cr    & 84     & 20         \\
$^{55}$Fe & 996       & $^{54}$Fe    & 5.8     &  $^{56}$Fe    & 92     & 16         \\
$^{75}$Se & 120       & $^{74}$Se    & 0.89    &  $^{76}$Se    & 9.4    & 11        \\
$^{85}$Sr & 65        & $^{84}$Sr    & 0.56    &  $^{86}$Sr    & 9.9    & 18         \\
$^{97}$Ru & 2.9       & $^{96}$Ru    & 5.5     &  $^{98}$Ru    & 1.9    & 0.3        \\
$^{103}$Pd& 17        & $^{102}$Pd   & 1.0     &  $^{104}$Pd   & 11.1   & 11          \\
$^{109}$Cd& 463       & $^{108}$Cd   & 0.89    &  $^{110}$Cd   & 12.5   & 14          \\
$^{113}$Sn& 115       & $^{112}$Sn   & 0.97    &  $^{114}$Sn   & 0.66   & 0.7         \\
$^{121}$Te& 16.8      & $^{120}$Te   & 0.09    &  $^{122}$Te   & 2.6    & 29         \\
$^{127}$Xe& 36        & $^{126}$Xe   & 0.09    &  $^{128}$Xe   & 1.9    & 21         \\
$^{133m}$Ba& 1.6      & $^{132}$Ba   & 0.1     &  $^{134}$Ba   & 2.4    & 24         \\
$^{139}$Ce& 138       & $^{138}$Ce   & 0.25    &  $^{140}$Ce   & 88     & 352          \\
$^{153}$Gd& 239       & $^{152}$Gd   & 0.2     &  $^{154}$Gd   & 2.2    & 11         \\
$^{159}$Dy& 144       & $^{158}$Dy   & 0.095   &  $^{160}$Dy   & 2.3    & 24        \\
$^{165}$Er& 0.43      & $^{164}$Er   & 1.6     &  $^{166}$Er   & 33.5   & 21        \\
$^{169}$Yb& 32        & $^{168}$Yb   & 0.13    &  $^{170}$Yb   & 3.04   & 23     \\
$^{175}$Hf& 70        & $^{174}$Hf   & 0.16    &  $^{176}$Hf   & 5.3    & 33     \\
$^{181}$W & 121       & $^{180}$W    & 0.12    &  $^{182}$W    & 26.5   & 221     \\
$^{191}$Pt& 2.8       & $^{190}$Pt   & 0.014   &  $^{192}$Pt   & 0.78   & 56     \\
$^{193m}$Pt& 4.3      & $^{192}$Pt   & 0.78    &  $^{194}$Pt   & 33     & 42      \\
\hline
\end{tabular}
\end{center}
\end{table}

\subsection{\boldmath \bf Radioisotopes via the ($\gamma$,p) 
reaction \unboldmath}

Even when excited beyond the proton binding energy, a nucleus does 
not necessarily lose a proton. The latter is bound by the Coulomb 
barrier, leading to a suppression of the proton loss channel.
Only for excitation well beyond the proton binding energy, the proton 
gains enough kinetic energy for tunneling efficiently through 
the Coulomb barrier. However, such excitation energies are usually also 
above the neutron binding energy or even the two-neutron binding energy. 
Hence neutron emission competes with proton emission and 
the cross sections for ($\gamma$,p) reactions may be one
order of magnitude lower than the competing channels (compare Fig.~\ref{fig3}).
Thus, the achievable specific activity (specific activity with respect 
to the target mass) is limited for ($\gamma$,p) reactions. However, 
the product isotope differs chemically from the target since it has 
one proton less ($Z_{\rm product} = Z_{\rm target} -1$). After irradiation 
a chemical separation of the product isotope from the target can be 
performed, ultimately resulting in a high specific activity that is 
only compromised by competing reactions leading to other isotopes
of the product element (such as ($\gamma$,np), 
($\gamma$,2n)EC/$\beta^+$, etc.) or product burn-up by ($\gamma$,n). 

\begin{enumerate}

\item $^{47}$Sc:\\ 
Besides the $^{48}$Ca($\gamma$,n)$^{47}$Ca$\rightarrow^{47}$Sc reaction, 
$^{47}$Sc can also be produced via the $^{48}$Ti($\gamma$,p) $^{47}$Sc 
reaction. The established Sc/Ti separation schemes can be employed 
for the chemical processing.
Compared to the $^{47}$Ti(n,p) way here the direct production of disturbing 
long-lived $^{46}$Sc (via $^{46}$Ti(n,p) or $^{47}$Ti($\gamma$,p), 
respectively) can be limited more easily, since $^{48}$Ti is the most 
abundant titanium isotope and can be enriched more easily to high abundance.
However, the irradiation times have to be kept relatively short to 
prevent excessive formation of $^{46}$Sc impurity by $^{47}$Sc($\gamma$,n) 
reactions.

\item $^{67}$Cu:\\ 
$^{67}$Cu is also a promising beta-emitter for targeted radiotherapy. 
Alike $^{47}$Sc it has a sufficiently long half-life for accumulation 
in the tumor cells when bound to antibodies and its 185~keV gamma ray 
allows imaging with SPECT or gamma cameras. Together with the PET imaging 
isotopes $^{61}$Cu or $^{64}$Cu it forms a ``matched pair''.
The usual production routes $^{68}$Zn(p,2p), $^{70}$Zn(p,$\alpha$) or 
$^{64}$Ni($\alpha$,p) are all characterized by low yields. The former 
requires energetic protons ($\gg 30$~MeV from larger cyclotrons)
and the latter two methods use expensive enriched targets with low 
natural abundances. The alternative production via $^{67}$Zn(n,p) requires 
a very high flux of fast neutrons, which is only available in few reactors.
At present clinical trials with $^{67}$Cu are hindered by insufficient 
supply \cite{Riv05}. Production via $^{68}$Zn($\gamma$,p) reactions induced 
by intense $\gamma$ beams provides higher activities 
and uses more abundant, and, hence cheaper $^{68}$Zn targets. 
The established Cu/Zn separation schemes can be employed for the 
required chemical processing.

\item Isotopes with higher $Z$:\\
In principle also heavier $\beta^-$ emitters used for radionuclide 
therapy such as $^{131}$I, $^{161}$Tb or $^{177}$Lu 
could be produced by ($\gamma$,p) reactions (on $^{132}$Xe, 
$^{162}$Dy or $^{178}$Hf targets respectively).
However, for higher $Z$ the increasing Coulomb barrier leads to 
small production cross sections. Production in high flux reactors by 
neutron-induced fission (for $^{131}$I) or by neutron capture 
on the ($A-1$,$Z-1$) target ($^{130}$Te, $^{160}$Gd or $^{176}$Yb 
respectively), followed by decay and chemical separation 
leads to products with excellent specific activity and is more economic.
  

\end{enumerate}

\begin{table}[t]
\caption{Therapy radioisotopes that can be produced in $(\gamma$,p) reactions}
\bigskip
\begin{center}
\begin{tabular}{cccc} \hline
isotope   & mean beta& $T_{1/2}$ & target isotope \\ 
          & energy  &            & natural abundance  \\
          &   (keV) &  (days)    & (\%)                          \\ \hline
$^{47}$Sc &  162    & 3.35  & $^{48}$Ti  (73.7\%)             \\      
$^{67}$Cu &  141    & 2.58  & $^{68}$Zn  (18.7\%)             \\
$^{131}$I &  182    & 8.03  & $^{132}$Xe (26.9\%)        \\
$^{161}$Tb & 154    & 6.91  & $^{162}$Dy (25.5\%)      \\
$^{177}$Lu & 134    & 6.65  & $^{178}$Hf (27.3\%)      \\
\hline
\end{tabular}
\end{center}
\end{table}

\subsection{\boldmath \bf Radioisotopes via the ($\gamma$,2n) reaction
             \unboldmath}

\begin{enumerate}

\item $^{44}$Sc:\\ 
$^{44}$Sc is a promising metallic PET tracer that emits a 1157~keV gamma-ray 
quasi-simultaneously with the positron. With a suitable detection system 
(Compton telescope plus PET camera), a triple coincidence (gamma rays 
of 511~keV, 511~keV, and 1157~keV) can be detected \cite{Gri07}. Hence, 
for each triple-event the point of emission is derived instead of the usual 
line-of-response, leading to improved position resolution at reduced dose 
to the patient. Moreover $^{44}$Sc forms a ``matched pair'' with $^{47}$Sc, 
a therapy isotope discussed above. $^{44}$Sc can be obtained from 
$^{44}$Ti/$^{44}$Sc generators where the parent isotope $^{44}$Ti is very 
long-lived ($T_{1/2}=60$ years). Despite the very favorable properties 
of $^{44}$Sc, this isotope is not yet used in clinical routine, since 
the generator isotope $^{44}$Ti is difficult to produce and therefore 
prohibitively expensive. The current cost for 200~MBq of $^{44}$Ti, the 
typical activity for one $^{44}$Ti/$^{44}$Sc generator for
human use, is about 2 million EUR! Exposing enriched $^{46}$Ti 
(natural abundance 8\%) to an intense $\gamma$ beam allows producing 
$^{44}$Ti by ($\gamma$,2n) reactions. It will take about three weeks of 
irradiation in a mid-sized $\gamma$ beam facility to generate 200~MBq of 
$^{44}$Ti, but such a generator (or several subsequent generators that use the 
same recycled $^{44}$Ti activity) can be eluted several times a day and 
serve for tens of years.



\item $^{224}$Ra/$^{212}$Pb/$^{212}$Bi:\\
Via $^{226}$Ra($\gamma$,2n) reactions the isotope $^{224}$Ra 
($T_{1/2}$ = 3.66 d) from the thorium chain can be
obtained, where the noble gas $^{220}$Rn isotope can be extracted easily. 
The $\alpha$ emitter $^{212}$Bi ($T_{1/2}$ = 60 min) in this decay chain 
or its mother isotope $^{212}$Pb are also considered for targeted 
alpha therapy, e.g. for malignant melanoma metastases \cite{Has01,Mia05}. 

\end{enumerate}

\subsection{\bf Other photonuclear reaction channels}
In ($\gamma$,2p) reactions even two protons must overcome the Coulomb 
barrier, making this reaction channel even less likely than the 
($\gamma$,p) reaction. Measured cross-sections for ($\gamma$,2p) reactions 
exist for $^{63}$Cu($\gamma$,2p)$^{61}$Co \cite{Ant95}. They range 
from few $\mu$-barn at 30~MeV to 14~$\mu$barn at 60~MeV.
Also for ($\gamma$,$\alpha$) reactions the higher Coulomb barrier leads 
to small cross-sections in the microbarn range.
Usually other production reactions provide better yields, making these types
of photonuclear reaction less competitive.

Photo-fission of uranium or thorium targets allows production of $^{99}$Mo 
and other isotopes with highest specific activity. 
However, the here proposed $\gamma$ beams with high flux density 
are not suitable since they lead to an excessive 
target heating. In each fission about 200~MeV are released, the dominant 
part as kinetic energy of the fission fragments that will stop 
within few $\mu$m range. This would lead to power densities of several 
hundred kW per cm$^3$, impossible to cool from the targets.


\section{Photonuclear Activation for Brachytherapy Applications}

Certain nuclear medicine applications use the radioisotopes ``directly'', 
i.e. not necessarily coupled to a bio\-mole\-cule.  

There are various applications for micro- or nanoparticles that are doped 
with radioisotopes. They can be used for radioembolization of ,e.g., 
liver cancer or liver metastases \cite{Gar10}. The radioactive 
microparticles are directly injected 
into the arteries supplying the tumor, then irradiate the latter 
with their medium-range radiation 
(beta particles, low-energy X-rays or gamma rays). 
Similarly, nanoparticles are considered for targeting tumors.
The radioisotopes can be introduced into the micro- or 
nanoparticles in various ways:
\begin{enumerate}
\item The radioisotopes can be added to the raw materials used in 
the chemical synthesis of the micro- or nanoparticles. However, 
this makes the processing much more involved, since radioactive material 
has to be handled and the respective radiological and contamination 
issues have to be addressed in the production facility.
\item The radioisotopes can be implanted in form of a radioactive ion 
  beam into the ready-made micro- or nanoparticles \cite{Rav06}. 
  This method is quite universal, allowing to dope even with radioisotopes 
  of elements that are usually not soluble in or chemically compatible with 
  the matrix. However, the radioactive isotopes first need to be brought 
  into a radioactive ion beam which may be more involved depending on the 
  chemical element.
\item A stable precursor of the radioisotope can be introduced prior to 
  the chemical synthesis of the micro- or nanoparticles or ion-implanted 
  after synthesis. Then the precursor is transmuted in a nuclear reaction 
  into the desired radioisotope. However, the micro- or nanoparticles may 
  be sensitive to radiation damage. Hence activation ,e.g., in a nuclear 
  reactor could damage them such that they are no longer usable in 
  in-vivo applications. For neutron activation it has been shown that
  resonance capture of epithermal neutrons (``adiabatic resonance crossing'' 
  method) can be of advantage to overcome this problem \cite{Abb09}. 
  Here we propose a complementary method of activation by photonuclear 
  reactions. For the isotopes listed in table \ref{slightly-n-def} the 
  general advantages discussed above apply. In addition the high 
  cross section ratio of ``useful'' photonuclear reactions 
  versus ``disturbing'' reactions causing radiation damage
  allows obtaining relatively high activities.
\end{enumerate}

Radioisotopes can also be bound in larger solid matrices that are 
then mechanically (surgically) introduced into the body 
or brought close to it to irradiate tumors or benign diseases. 
Such a so-called brachytherapy is today routinely used to treat 
prostate cancer by permanently introduced seeds containing 
radioactive $^{125}$I \cite{Aal09}. 
It is also useful to prevent in-stent restenosis by intravascular 
brachytherapy using radioactive stents \cite{Ens05}, prevent closure 
of the pressure relief channel in glaucoma filtering surgery by 
radioactive implants \cite{Ass07} or perform other anti-inflammatory or 
anti-proliferative treatments. Photonuclear reactions could simplify 
the production of the respective stents or seeds. Instead of introducing 
the radioactive isotopes in the production process or ion-implanting 
it afterwards, it will be possible to produce the stents or seeds in 
their final form and then activate a previously included stable precursor 
isotope by photonuclear reactions. Selective photonuclear reactions assure 
to keep the radiation damage of the matrix low and avoid an 
unwanted production of disturbing radioisotopes by activation of the matrix.

\section{Advantages of the Proposed Photonuclear Reactions 
            over Existing Technologies}
The intense brilliant $\gamma$ beam will allow to produce radioisotopes 
with rather high specific activity. Advantages of $\gamma$ ray beams with 
small opening angle are:

\begin{itemize}

\item The produced radioisotopes are concentrated in a small 
   target volume, hence resulting in much higher specific activity than usual.

\item Much less of the (often costly) target material is required.

\item Radioactive targets are more efficiently converted into the required
  product isotopes, hence more compact and less active targets can be
  employed, resulting in less activity to be handled and less dose 
  rate.
\end{itemize}

 Additional advantages when using the low bandwidth $\gamma$ ray beams are:
\begin{itemize}
\item The higher cross section for monochromatic beams leads to a short
interaction 
  length (cm or less). This leads to an additional reduction of the
  required target mass, hence further reducing the target costs and increasing
  the specific activity.

\item Compared to Bremsstrahlung beams a much reduced $\gamma$ ray heating 
per useful reaction rate occurs since the $\gamma$ rays in the useful 
energy range are not accompanied by an intense low-energy tail. 
Moreover the usual equilibrium between $\gamma$-rays and electrons 
(which are responsible for the actual heating)
will build up only for very thick targets. 

\item Much reduced radiation damage due to quasi-mono\-chromatic beams
will make it possible to first dope and then activate materials (e.g. organic, 
nanoscale,\dots ) that would not withstand irradiation in a nuclear 
reactor or a brems\-strahlung $\gamma$ ray spectrum.

\item Isotopic enrichment may not necessarily be needed, when for a 
given $\gamma$ energy the wanted cross section is much higher than 
for other isotopes. In particular, the fine structure
of the Pygmy dipole resonance (PDR), probably similar to the 
giant dipole resonance (GDR), could be exploited.
  
\item Also less stringent requirements exist concerning isotopic enrichment 
or chemical impurities of the target materials if the $\gamma$ ray energy
is chosen such that the maximum cross sections of the wanted production 
channels corresponds to minima in the cross section of activation of 
impurities. 

\item Selective production reduces the overall activity level of the irradiated
  target and reduces the challenge to the chemical post-processing.

\end{itemize}

Moreover, there are practical advantages of photonuclear reactions compared to 
charged-particle induced reactions: Radioactive targets like $^{226}$Ra or 
targets that risk to react heavily in contact with cooling water (e.g. alkali
metals) can be safely encapsulated into relatively thick metal walls, 
since gamma rays penetrate easily and cause less heating of the walls 
than charged particles do.

A further optional increase of the specific activity is possible by:

\begin{itemize}
\item Enriched target isotopes may be used. 

\item A thin target or a stack of thin target foils interleaved with 
     a different solid, liquid or gas that acts as catcher of recoil ions. 
   Extraction and separation of the recoiled product isotopes can be 
  performed with the usual radiochemical methods.

\item If the produced radioisotope belongs to a different chemical element than
the target (e.g. for ($\gamma$,p) reactions), a usual radiochemical
post-processing (e.g. ion exchange chromatography, liquid-liquid extraction, 
etc.) can be employed to separate the product element from remainders of 
the target element and thus increase the specific activity of the product.

\item A product isotope that decays to a radioactive daughter isotope with
medical applications allows producing a generator.

\end{itemize}


\section*{Conclusion and Outlook}

Laser beams revolutionized atomic physics and its applications, not only due 
to their coherence, but often just due to their high photon flux density or 
their high spectral photon flux density. Similarly gamma beams open many new 
possibilities in nuclear physics and its applications.
Alike in atomic physics such beams can be used to pump a good fraction 
of the nuclear ground state population via excited levels into an isomeric 
state. While (far) ultraviolet laser beams can efficiently 
photoionize atoms by exciting a valence electron to an unbound state, 
energetic gamma beams can efficiently excite a nucleon (neutron or proton) 
into an unbound state leading to photo-dissociation and creation
of a new isotope. When these reactions are resonantly enhanced the 
monochromaticity of laser or gamma beams respectively is decisive.

Using the new $\gamma$ beam facilities we can use compact targets, 
which are exposed to the gamma radiation and undergo photonuclear 
reactions such as ($\gamma,\gamma')$, ($\gamma$,n), 
($\gamma$,p), etc. to form radioisotopes. After a suitable irradiation 
time, a radioisotope with high specific activity is produced. After the 
usual radiochemical and radiopharmaceutical 
steps (such as optionally dissolving of the target, optionally chemical 
purification, labeling, quality control,\dots) a 
radiopharmaceutical product is created for use in diagnostic or 
therapeutic nuclear medicine procedures. The produced radioisotope 
may be used directly for nuclear medicine applications.

The investment and running costs of the proposed $\gamma$ beam 
facilities are of the order of 40~MEUR and few MEUR/year. This is 
cheaper than a high-flux reactor
, but more expensive than compact cyclotrons that provide charged 
particle beams with 10 to 20 MeV energy
suitable for production of PET tracers. World-wide 
more than 600 such cyclotrons exist, often based at hospitals or close-by. 
They provide regularly the short-lived PET isotopes $^{18}$F ($T_{1/2} 
= 110$~min), $^{11}$C (20 min), $^{13}$N (10 min) and $^{15}$O (2 min) for 
molecular imaging applications. Although it would be possible to produce 
also such isotopes by photonuclear reactions 
(e.g. $^{20}$Ne($\gamma$,np)$^{18}$F), a more complex Compton backscattering
facility would be clearly an overkill for such applications.

The selection of radioisotopes presently used in nuclear medicine routine 
applications is actually dictated by the radioisotope's availability 
at reasonable costs. This selection is not necessarily optimized for 
all clinical applications with respect to optimum nuclear 
(half-life, decay radiation) and chemical properties 
(labeling efficiency, in-vivo stability of compounds, etc.). 
Other radioisotopes might be advantageous for the patient but 
are presently not available. The main advantage of the $\gamma$ beam 
facility is the new and rather unique access to radioisotopes or
isomers with high specific activity that can complement and extend 
the choice of radioisotopes for nuclear medicine applications.


\vspace{5mm}

{\bf Acknowledgement}\\

We acknowledge helpful discussions with C.~Barty and R.~Hajima. 
We enjoyed the close collaboration with V.~Zamfir, who is heading
ELI-NP. We were supported by the DFG Clusters of Excellence:
Munich Centre for Advanced Photonics (MAP) and UNIVERSE.

\end{document}